\documentclass[%
 reprint,
nofootinbib,
 amsmath,amssymb,
 aps,
]{revtex4-2}
\usepackage{graphicx, color}
\usepackage{amsmath}
\usepackage{ulem}
\usepackage{slashed}
\usepackage{float}
\usepackage{dcolumn}
\usepackage{bm}

\newcommand\simgreater{\buildrel > \over \sim}

\newcommand{\msun}{\mbox{$\mathrm{M}_{\odot}\;$}}

\newcommand{\be}{\begin{equation}}
\newcommand{\ee}{\end{equation}}
\newcommand{\beq}{\begin{eqnarray}}
\newcommand{\eeq}{\end{eqnarray}}
\newcommand{\bv}{Brunt-V\"ais\"al\"a}
\newcommand{\gm}{$g$-mode}
\newcommand{\gms}{$g$-modes}

\begin{document}

\newcommand{\prash}[1]{\textcolor{red}{#1}}
\newcommand{\prak}[1]{\textcolor{blue}{#1}}
\newcommand{\alex}[1]{\textcolor{magenta}{#1}}
\newcommand{\cons}[1]{\textcolor{purple}{#1}}

\preprint{APS/123-QED}

\title{g-mode oscillations in hybrid stars:\\A tale of two sounds}

\author{Prashanth Jaikumar}
\email{prashanth.jaikumar@csulb.edu}
\affiliation{Department of Physics and Astronomy, California State University Long Beach, Long Beach, California 90840, USA
}%
\author{Alexandra Semposki}
\email{as727414@ohio.edu}
\author{Madappa Prakash}%
 \email{prakash@ohio.edu}
\affiliation{
 Department of Physics and Astronomy, Ohio University, Athens, Ohio 45701, USA
}%
\author{Constantinos Constantinou}
\email{cconstantinou@ectstar.eu}
\affiliation{INFN-TIFPA, Trento Institute of Fundamental Physics and Applications, Povo, 38123 Trento, Italy}
\affiliation{European Centre for Theoretical Studies in Nuclear Physics and Related Areas, Villazzano, 38123 Trento, Italy}

\date{\today}
\begin{abstract}
 We study the principal core \gm~oscillation in hybrid stars containing quark matter and find that they have an unusually large frequency range  ($\approx$ 200 - 600 Hz) compared to ordinary neutron stars or self-bound quark stars of the same mass. Theoretical arguments and numerical calculations that trace this effect to the difference in the behaviour of the equilibrium and adiabatic sound speeds in the mixed phase of quarks and nucleons are provided. We propose that the sensitivity of core \gm~oscillations to non-nucleonic matter in neutron stars could be due to the presence of a mixed quark-nucleon phase. Based on our analysis, we conclude that for binary
 mergers where one or both components may be a hybrid star, the fraction of tidal energy pumped into resonant \gms~in hybrid stars can exceed that of a normal neutron star by a factor of 2-3, although resonance occurs during the last stages of inspiral. A self-bound star, on the other hand, has a much weaker tidal overlap with the \gm. The cumulative tidal phase error in hybrid stars, $\Delta\phi\cong$ 0.5 rad, is comparable to that from tides in ordinary neutron stars, presenting a challenge in distinguishing between the two cases. However, should the principal \gm~be excited to sufficient amplitude for detection in a postmerger remnant with quark matter in its interior, its frequency would be a possible indication for the existence of non-nucleonic matter in neutron stars. \\

\end{abstract} 

\maketitle


\section{\label{sec:level1}Introduction}
The core of a neutron star (NS) can, in principle, support phases of
dense deconfined quark matter (QM)~\cite{1975PhRvL..34.1353C}. 
Confirmation of the presence of quarks in NSs, however, has not been possible either through observations or from lattice-gauge calculations of finite baryon density matter starting from the Lagrangian of Quantum Chromodynamics (QCD). Although perturbative calculations of QM have been performed~\cite{Kurk1,Kurk2,Kurk3}, their applicability is limited to baryon densities $n_B \gtrsim 40n_s$~\cite{2020NatPh..16..907A}, where $n_s\simeq 0.16~\rm{fm^{-3}}$ is the nuclear matter saturation density. Such densities, however, lie well beyond the central densities $n_c$ in the range 3-8$n_s$ of observed NSs.   
In view of this conundrum, theoretical studies of QM in NSs have been exploratory in nature by positing either a sharp 1st-order or a smooth crossover hadron-to-quark phase transition. Depending on the treatment of the phase transition and the  equations of state (EOSs) of hadronic and quark matter, either a phase of pure QM or a phase in which hadrons are admixed with quarks can be realized (for a detailed account, see Ref.~\citep{2019PhRvD.100j3022H} and an extensive list of references therein). In either case, 
stars with quarks are difficult to distinguish from normal 
NSs based on the knowledge of masses and radii alone as similar results can be obtained with both. While the long-term cooling of a NS
can be affected by the presence of quarks, cooling data are relatively sparse and gathered over decades~\citep{2013arXiv1302.6626P,Potekhin:2015qsa}. 
Gravitational wave observations from compact binary mergers can be another probe of the EOS, but currently, constraints on tidal polarizability~\cite{Hind,Damour,Post} 
 from gravitational wave data~\cite{2017PhRvL.119p1101A} 
are consistent with both normal and quark-containing stars, depending on the theoretical assumptions made~\citep{2019PhRvD.100j3022H,2020arXiv200914274P,2019EPJA...55...97T}. 

In this paper, we are particularly interested in how NS oscillations can shed light on the presence of QM in stars that contain an admixture of nucleons and quarks (termed hybrid stars). Andersson and Kokkotas~\citep{AK2001} have proposed that NS oscillations (in particular, the $f,p$ modes) could be a ``fingerprint" for the supra-nuclear EOS in gravitational wave data. A review of potential signatures of QM in NSs in the multi-messenger era, including the role of their oscillations, can be found in~\cite{2019JPhG...46k4001A}.
Along these lines, 
we offer in this work a new 
diagnostic of deconfined QM in NSs based on asteroseismology. We show that a steep rise in the frequency of
the principal \gm~(gravity mode) occurs as soon as QM appears in a mixed phase in the core, exceeding the typical core \gm~frequency of a nucleonic star by a factor of two or more. This rise is essentially driven by a drop in the local equilibrium speed of sound at the onset of the phase transition, while the adiabatic sound speed changes only slightly. If this \gm~becomes resonant with the tidal force during the late stages of a binary inspiral, the resulting energy transfer from the orbital motion to the star via tidal coupling can affect the phase of the gravitational waveform, and potentially signal a hybrid star.

\indent NS oscillations are categorized by the nature of the dominant restoring force for the perturbation in question. Several types of modes can be supported by a star and it is desirable to investigate as many of them as possible in detail. These modes are typically excited and sustained in different regions of the star and their amplitudes and damping rates are subject to considerable uncertainty. Here, for reasons that will become apparent, we focus our attention on the \gm~and its coupling to gravitational waves.

A \gm~is a specific kind of non-radial fluid oscillation
initiated when a parcel of fluid is displaced against the background of a stratified environment~\citep{RG,RG2}. While pressure equilibrium is rapidly restored via sound waves, chemical equilibrium can take longer causing buoyancy forces to oppose the displacement. Since cold NSs are not convective, the opposing force sets up stable oscillations throughout the core, with a typical frequency, called the (local) \bv\, frequency~\cite{Cox}, which depends on the difference between the equilibrium and adiabatic sound speeds as well as the local metric coefficients. Convectively stable \gms~exist for a wide range of 
models of the EOS~\citep{Lai94}. Though the \gm~in NSs has been studied before~\citep{Finn87,Stroh93,McD90}, with recent works incorporating additional features like hyperonic matter and superfluidity~\citep{ULee,Prix_2002,Comer,GK13, Gualtieri,Kantor14,Dommes15,Pass16,Yu17,Yu-Wein,Rau18}, the novel aspect of our work is the investigation of the \gm~frequency in a phase transition from nuclear matter to a mixed phase of quarks and nucleons. We point out that a similar result was obtained in~\citep{Dommes15} for superfluid hyperonic stars. However, the calculations presented there were 
for a fixed 
NS mass of about 1.64$M_\odot$ 
with a radius of 13.6 km. While their chosen hyperonic EOS does support a maximum mass of $2.015M_\odot$~\cite{Gusakov14},
the nuclear and quark EOSs chosen in our work satisfy additional observational  and experimental constraints, and presents the effect for a wide range of masses up to the observed maximum.

This paper also extends the results of~\citep{Wei_2020} by incorporating aspects of General Relativity in the fluid oscillation equations (while remaining within the Cowling approximation), updating the nuclear EOS to include consistency with radius constraints from tidal polarizability~\citep{AbbottEOS} and {\it NICER} data~\citep{Riley-nicer,Miller-nicer} on the radius of $\simeq 1.4 M_\odot$ NS. We also 
provide new analytical results for the two sound speeds in a mixed phase of quarks and nucleons.
Our study can be of practical interest to investigations of the sound speed in NSs, which is attracting renewed attention~\cite{2015PhRvL.114c1103B,2020PhRvC.102e5801K,2019PhRvL.122l2701M,zl2020}. The matter of detecting \gms~from the pre- or post-coalescence phase of binary NS mergers is not addressed in detail here, but we present an estimate of its impact on the accumulated tidal phase up to the merger. 

\indent It is pertinent to note that oscillation modes other than \gms~can also potentially be affected by the presence of QM in NSs. Radial oscillation modes, which however do not couple to gravitational waves, were studied in~\cite{Brillante}. Among non-radial modes, the $i$-mode (interface mode) has been recently investigated for the special case of crystalline QM surrounded by a hadronic envelope in~\cite{2020arXiv201213000L} and its frequency can range from (300-1500) Hz, which can be probed by current or third generation interferometric detectors. The $r$-mode (Rossby mode) frequency and its damping rate for NSs containing QM also differs from a purely nucleonic one~\cite{Jaikumar_2008,2008arXiv0806.3359S,2010PhRvC..82e5806R}. The $s$-mode (shear mode) can be excited in a mixed phase of quarks and nucleons and is sensitive to the shear modulus of structures in the mixed phase~\cite{Sotani12}, probing the surface tension of QM. The \gm~oscillation in stars containing QM has been studied in the case of a sharp interface~\cite{Flores13,Ranea_Sandoval_2018} between hadronic and quark matter, yielding the spectrum of so-called discontinuity
\gms, but these works assume a strong 1st-order phase transition and a large value of the surface tension for QM, while we study the case of a mixed phase of significant extent that would be favored if the same surface tension 
were small enough\footnote{Here, we do not explicitly study surface and curvature effects or the impact of a non-trivially structured mixed phase on the oscillation spectrum~\cite{1995sqm..conf..298H,Sotani12,2018PhRvD..98c4031X}, but a more complete treatment of \gm s in hybrid stars should address these issues.}. The $g$-mode for baryonic stars with superfluid effects was studied in~\citep{ULee,Prix_2002,Comer,GK13, Gualtieri,Kantor14,Dommes15,Pass16,Yu17,Yu-Wein,Rau18} highlighting the subtle role of temperature and composition gradients in driving the \gm. In this work, we 
investigate the composition gradients induced by the mixed phase of quarks and nucleons, which supports unusually high-frequency \gms~through its effect on the adiabatic and equilibrium sound speeds. 

The organization of this paper is as follows. In Sec.~\ref{sec2}, we introduce the governing equations of the \gm~and outline its relation to the two sound speeds. In Sec.~\ref{sec3}, we present the EOS models in the nucleonic and quark sectors chosen for our study of the \gm~spectrum. The rationale for our parameter choices and the basic features of these models are highlighted here for orientation. We stress that our choices are representative, but not exhaustive. In Sec.~\ref{sec4}, we derive expressions for the two sound speeds in the mixed phase of quarks and nucleons. Results for the sound speeds, the \bv\, frequency and the \gm~frequency in nucleonic, quark and hybrid stars are gathered and interpreted in Sec.~\ref{sec5}. The tidal forcing and phase error due to 
\gm~excitation are estimated in Sec.~\ref{sec6}. Our summary, conclusions and outlook are contained in Sec.~\ref{sec7}. The appendices contain details about the determination of parameters in the nuclear EOS model, the resulting NS structural properties and the two sound speeds.

\section{$g$-mode Oscillations} 
\label{sec2}
In this section, we outline the equations for fluid oscillations and non-rotating NS structure that were used to determine the eigenfrequencies of the $g$-mode. In general, the oscillatory displacement of a fluid element in a spherically symmetric star is represented by a vector field ${\vec{\xi}}^{nlm}(\vec{r}){\rm e}^{-i\omega t}$ with $n,l$ and $m$ denoting the radial, azimuthal and magnetic mode indices. To be precise, the frequencies also carry subscripts $nlm$  
implicitly understood, with degeneracies that are broken in more realistic cases such as with rotation or magnetic fields. For even-parity or spheroidal modes, separation into radial and tangential components yields $\xi_r^{nlm}(\vec{r})$ = $\eta_r^{nl}(r)Y_{lm}(\theta,\phi)$ and $\xi_{\perp}^{nlm}(\vec{r})$ = $r\eta_{\perp}^{nl}(r)\nabla_{\perp}Y_{lm}(\theta,\phi)$, respectively, where $Y_{lm}(\theta,\phi)$ are the spherical harmonics. From the perturbed continuity equation for the fluid, the tangential function $\eta_{\perp}$ can be traded for fluid variables as $\delta p/\epsilon$ = $\omega^2r\eta_{\perp}(r)Y_{lm}(\theta,\phi){\rm e}^{-i\omega t}$, where  $\delta p$ is the corresponding local (Eulerian) pressure perturbation and $\epsilon$ the local energy density. Within the relativistic Cowling approximation\footnote{The Cowling approximation neglects the back reaction of the gravitational potential and reduces the number of equations we have to solve. While this approximation is not strictly consistent with our fully general relativistic (GR) treatment of the equilibrium structure of the star, it does not change our conclusions qualitatively or even quantitatively, since this approximation is accurate for $g$-mode frequencies at the few \% level \citep{Grig}.}, the equations of motion to be solved to determine the frequency of a particular mode are~\citep{McD83,RG,Kantor14}
\begin{align}
-\frac{1}{\mathrm{e}^{\lambda / 2} r^{2}} \frac{\partial}{\partial r}\left[\mathrm{e}^{\lambda / 2} r^{2} \xi_{r}\right]+\frac{l(l+1) \mathrm{e}^{\nu}}{r^{2} \omega^{2}} \frac{\delta p}{p+\epsilon} 
-\frac{\Delta p}{\gamma p} = 0  \nonumber \\ 
\frac{\partial \delta p}{\partial r}+g\left(1+\frac{1}{c_{\mathrm{s}}^{2}}\right) \delta p+\mathrm{e}^{\lambda-\nu} h
\left(N^{2}-\omega^{2}\right) \xi_{r} = 0 \,, 
\label{oscr}
\end{align}
where $h=p+\epsilon$, and we have suppressed the indices on $\omega$ and $\xi$. 
Equation~(\ref{oscr}) involves thermodynamic quantities that follow from the specific EOS. Specifically, $p$ denotes pressure, $\epsilon$ energy density, and $\gamma$ the adiabatic index of the fluid. The Lagrangian variation of the pressure enters as $\Delta p$, and is related to the Eulerian variation $\delta p$ through the operator relation $\Delta \equiv \delta + \xi\cdot\nabla$. The symbol $c_s$ denotes the adiabatic sound speed, which is related to the adiabatic index  as $c_s^2=\gamma p/(\mu_n n_B)$ where $\mu_n$ is the neutron chemical potential\footnote{In beta-equilibrated charge neutral matter, the neutron chemical potential is sufficient to determine all other chemical potentials.} and $n_B$ the local baryon density. The equilibrium sound speed enters through the \bv\, frequency ($N$) which is given by 
\beq
\label{bvf}
    N^{2}\equiv g^{2}\Big ( \frac{1}{c_{e}^{2}}-\frac{1}{c_{s}^{2}} \Big ){\rm e}^{\nu-\lambda} \,,
\eeq
where $g=-\nabla\phi=-\nabla p/h$ with $h=\epsilon+p$ the enthalpy of the fluid. Finally, $\nu(r)$ and $\lambda(r)$ are metric functions of the unperturbed star which 
features in
the Schwarzschild {\it interior} metric ($r<R$):
\begin{eqnarray}
-\mathrm{d} s^{2} \equiv\, g_{\alpha \beta} \mathrm{d} x^{\alpha} \mathrm{d} x^{\beta}= &-\mathrm{e}^{\nu(r)} \mathrm{d} t^{2}+\mathrm{e}^{\lambda(r)} \mathrm{d} r^{2} \nonumber \\
&+r^{2}\left(\mathrm{d} \theta^{2}+\sin ^{2} \theta \mathrm{d} \varphi^{2}\right).
\label{Schwarz}
\end{eqnarray}
Explicitly,
\beq
e^{\lambda(r)}=\frac{1}{1-\left(\frac{2 G m(r)}{c^{2} r}\right)}
\eeq
and 
\begin{align}
e^{\nu(r)} = \exp \bigg[-\frac{2 G}{c^{2}} \int_{0}^{r}\left(\frac{\left(m(r^{\prime})+\frac{4 \pi p(r^{\prime}) r^{\prime 3}}{c^{2}}\right)}{r^{\prime}\left(r^{\prime}-\frac{2 m(r^{\prime}) G}{c^{2}}\right)}\right) d r^{\prime}\bigg]{\rm e}^{\nu_0},
\end{align}
where $m(r^\prime)$ is the enclosed mass of the star at $r^\prime$.
These metric functions must match to their exterior values at the surface 
$r=R$, hence the constant factor ${\rm e}^{\nu_0}$~\citep{chandrasekhar_non-radial_1991}.

In this work, we study the fundamental $g$-mode with $n$ = 1 and fix the mode's multipolarity at $l$ = 2. For the non-rotating stars we consider here, solutions are degenerate in $m$. Note that our definition of the ``fundamental" mode refers to the lowest radial order of the $g$-mode which also has the highest frequency. This should not be confused with the qualitatively different $f$-mode which is also referred to sometimes as the fundamental mode. Furthermore, overtones with lower frequency exist, but we do not perform any 
computations with them here, since the fundamental $g$-mode has the highest frequency and will be excited 
during the final stage of
the pre-merger phase when tidal forces are strongest. The system of equations in Eq.~(\ref{oscr}) cannot be solved analytically even with a simple model of a neutron star. Our aim will be to solve this numerically as an eigenvalue system for the $g$-mode frequency $\omega$. Physically, the solution to this system of equations, under the boundary conditions $\Delta p=0$ at the surface and $\xi_r,\,\delta p/\epsilon$ regular at the center, only exists for discrete values of the mode frequency $\omega$. These values represent the $g$-mode spectrum for a chosen stellar model. 
Because we have employed the Cowling approximation and ignored the perturbations of the metric that must accompany fluid perturbations, we cannot compute the imaginary part of the eigenfrequency (damping time) of the $g$-mode\footnote{The damping time of $g$-modes due to viscosity and gravitational wave emission, crudely estimated  in~\cite{Lai2,Wei_2020}, suggests that the \gm~can become secularly unstable for temperatures $10^8~{\rm K}<T<10^9~{\rm K}$ for rotational speeds exceeding twice the \gm~frequency of a static star.}. We turn now to discuss the EOS models for nucleonic and quark matter employed in this work.

\section{Models for the Equation of State}
\label{sec3}
The EOS models chosen in this work were predicated on the requirement that the squared sound speeds $c_e^2$ (see Eq.(\ref{css-pure})) and $c_s^2$ could be calculated straightforwardly. In the nucleonic sector, we employ the model of Zhao and Lattimer (ZL)~\cite{zl2020} which is consistent with nuclear systematics at and below the nuclear saturation density $n_s$. With suitable choices of the slope of the nuclear symmetry energy at $n_s$ (see below), this EOS is also consistent with the recent chiral effective theory calculations of Drischler et al.~\cite{Drischler} in which uncertainty estimates of the EOS up to $2n_s$ were provided (see Fig. 2 of this reference). In addition, the ZL EOS is able to support $\simeq 2M_\odot$ stars required by mass measurements of heavy NSs~\cite{twomass}, and is consistent with the recent radius measurements of $\sim 1.4M_\odot$ stars~\cite{Riley-nicer,Miller-nicer} and the tidal deformability estimates from the binary neutron star merger GW170817~\cite{AbbottEOS}.

Among the many models and treatments available in the quark sector~\cite{2019PhRvD.100j3022H}, we utilize the vMIT model of Gomes et al.~\cite{Gomes_2019} as a caricature of strongly interacting quarks at the densities attained within NSs.
Such interactions between quarks are required to satisfy astrophysical data, particularly those of heavy mass NSs. For the treatment of the nucleon-to-quark transition at supra-nuclear densities, we employ the Gibbs construction~\cite{Glen} 
which renders the transition to be smooth. Alternative models and treatments that feature strong first- or second-order phase transitions will be undertaken in subsequent work.

\subsection{The ZL EOS for Nucleonic Matter}
For completeness and to set the stage for the calculation of the two sound speeds in the next section, relevant details of the ZL model are provided below. 
The total energy density of interacting nucleons in neutron star matter (NSM) is 
\begin{align}
\epsilon_B = &\sum_{i=n,p} \frac{1}{\pi^2} \int_0^{k_{Fi}} k^2 \sqrt{M_B^2 + k^2} \,  dk \nonumber \\
&+ n_B V(n_n,n_p)\,,
\label{epsB}
\end{align}
where the Fermi momenta $k_{Fi} = (3\pi^2n_i)^{1/3}$ with $i=n,p$ and $n_B=n_n+n_p$, and $M_B$ is the baryon mass in vacuum. In the ZL model, interactions between nucleons are written as
%
\begin{align}
V(n_n,n_p) \equiv V(u,x) =&\,4x(1-x) (a_0 u + b_0 u^\gamma) \nonumber \\
&+ (1-2x)^2  (a_1 u + b_1 u^{\gamma_1}),
\end{align}
where $u=n_B/n_s$ and the proton fraction $x=n_p/n_B$. Adding and subtracting $a_0 u + b_0 u^\gamma$, the above equation can be rewritten as 
\begin{align}
\label{ZLinter}
V(u,x) &= V_0 + S_{2i}(u) (1-2x)^2 \quad \nonumber \\
\intertext{\rm with} V_0 &= a_0 u + b_0 u^\gamma, \quad \nonumber \\
\quad S_{2i}(u) &= (a_1 - a_0) u + b_1 u^{\gamma_1} - b_0 u^{\gamma}\,,
\end{align}
where the subscript ``$2i$'' in 
$S_{2i}$ refers to the interacting part of the total symmetry energy $S_2 = S_{2k} + S_{2i}$, with $S_{2k}$ representing the kinetic part.  
Expanding the kinetic part in Eq.~(\ref{epsB}) to order $(1-2x)^2$, we obtain the result\footnote{For the derivation of the kinetic part of the symmetry energy and its derivatives, see the Appendix.}
\beq
S_{2k} = \frac {1}{8} \left[ \frac {1}{n} \frac {\partial^2 \epsilon_{Bk}} {\partial x^2} \right]_{x=\frac{1}{2}}  = \frac{k_F^2}{6E_F} \,,
\eeq
where $k_F = (3\pi^2n_B/2)^{1/3}$ is the Fermi wave number of symmetric nuclear matter (SNM) and $E_F = \sqrt{k_F^2+M_B^2}$. Collecting the results, the energy per baryon relative to $M_B$ is given by
\begin{align}
\frac {\epsilon_B} {n_B} - M_B &=& E(u,x) = E_{{\rm SNM}} + S_2(u) (1-2x)^2 \nonumber \\
\intertext{\rm where} E_{{\rm SNM}} &=& T_{1/2} + V_{1/2} = T_{1/2} + (a_0 u + b_0 u^\gamma),  \nonumber \\
S_2(u) &=&  \frac{k_F^2}{6E_F} + 
(a_1 - a_0) u + b_1 u^{\gamma_1} - b_0 u^{\gamma} 
\,.
\label{asym}
\end{align}

The kinetic energy  per baryon $T_{1/2}$ in SNM ($x=1/2$) is given by the expression 
\begin{eqnarray}  
 T_{1/2} &=& \frac{ \epsilon_{1/2}^{\rm kin} }{n_B} - M_B \quad {\rm with} \nonumber \\
 \epsilon_{1/2}^{\rm kin} &=& \frac {2}{4\pi^2} \bigg[ k_F E_F \left(  k_F^2 + \frac{M_B^2}{2} \right) \nonumber \\
 &-& \frac 12 M_B^4 \ln \left( \frac {k_F+E_F}{M_B} \right)  \bigg] \,,
\end{eqnarray}
where $n_B$, $k_F$ and $E_F$ refer to those in SNM.

The  baryon pressure $p_B$ is
\beq
p_B = n_s u^2 \frac{dE_B}{du} = p_{\rm SNM} + n_s u^2 (1 - 2x)^2 \frac{dS_{2}(u)}{du} \,,
\eeq
where 
\begin{eqnarray}
p_{SNM} &=& p_{1/2}^{\rm kin} + n_s~( a_0 u^2 + \gamma b_0 u^{\gamma + 1}), \quad {\rm with} \nonumber \\
p_{1/2}^{\rm kin} &=&
\frac {2}{12\pi^2} \bigg[ k_F E_F \left(  k_F^2 - \frac 32 M_B^2 \right) \nonumber \\
&+& \frac 32 M_B^4 \ln \left( \frac {k_F+E_F}{M_B} \right)  \bigg], \quad {\rm and} \nonumber \\ 
u \frac{dS_{2}(u)}{du} &=& 
\frac 23 S_{2k}\left[1-18\left(\frac {S_{2k}}{k_F}\right)^2 \right] 
+ (a_{1} - a_{0}) u \nonumber \\
&+& b_{1}\gamma_{1}u^{\gamma_{1}} - b_{0}\gamma u^{\gamma}.
\end{eqnarray}
The incompressibility  $K_B$ in SNM is obtained from 
\beq
K_B &=& 9 \frac {dp_B}{dn_B} = 9 \left[ 2u \frac{dE_B}{du} + u^2\frac {d^2 E_B}{du^2} \right] \nonumber \\ 
&=& 9 \left\{ \frac {k_F^2}{3E_F} + \left[ 2a_0 u + \gamma (\gamma+1) b_0 u^\gamma \right] \right\}
\eeq

The energy per baryon in pure neutron matter (PNM in which $x=0$) relative to the baryon mass is
\begin{eqnarray}
E_{{\rm PNM}} &=&  T_0 + V_0 = T_0 + (a_1 u + b_1 u^\gamma_1) \nonumber \\
 T_0 &=& \frac{ \epsilon_0^{\rm kin} }{n_B} - M_B \quad {\rm with} \nonumber \\
 \epsilon_0^{\rm kin} &=& \frac {1}{4\pi^2} \bigg[ k_{Fn} E_{Fn} \left(  k_{Fn}^2 + \frac{M_B^2}{2} \right) \nonumber \\
 &-& \frac 12 M_B^4 \ln \left( \frac {k_{Fn}+E_{Fn}}{M_B} \right)  \bigg] \,,
 \label{EPNM}
\end{eqnarray}
where now $n_B=n_n$, $k_{Fn} = (3\pi^2 n_n)^{1/3} = 2^{1/3} k_F$, and $E_{Fn} = \sqrt{k_{Fn}^2+M_B^2}$.

The determination of the EOS constants in SNM and PNM, and relevant NS structural properties are summarized in Appendix A.

\subsection{The vMIT Equation of State for Quark Matter}
\label{vMIT-EOS}

In recent years, variations of the original bag model~\cite{1976PhLB...62..241B} have been adopted~\cite{Gomes_2019, Klahn1} to calculate the structure of NSs with quarks in their cores to account for $\geq 2 \mathrm{M}_{\odot}$ maximum-mass stars. Termed as vMIT or vBag models, the QCD perturbative results are dropped and replaced by repulsive vector interactions between quarks in such works. We will provide some numerical examples of the vMIT model for contrast with other models as those of the vBag model turn out to be qualitatively similar.

The Lagrangian density of the vMIT bag model is
\begin{equation}
\mathcal{L}=\sum_{i}\left[\bar{\psi}_{i}\left(i \slashed{\partial} -m_{i}-B\right) \psi_{i}+\mathcal{L}_{\mathrm{int}}\right] \Theta,
\end{equation}
where $\mathcal{L}_{\mathrm{int}} = \mathcal{L}_{\mathrm{pert}} + \mathcal{L}_{\mathrm{vec}}$ describes quarks of mass $m_{i}$ confined within a bag as denoted by the $\Theta$ function. For three flavors $i=u, d, s$ and three colors $N_{c}=3$ of quarks, the number and baryon densities, energy density, pressure and chemical potentials in the bag model are given by 
\begin{align} 
n_{i} &=2 N_{c} \int^{k_{F i}} \frac{d^{3} k}{(2 \pi)^{3}}, \quad n_{B}=\frac{1}{3} \sum_{i} n_{i} \\ \epsilon_{Q} &=2 N_{c} \sum_{i} \int^{k_{F i}} \frac{d^{3} k}{(2 \pi)^{3}} \sqrt{k^{2}+m_{i}^{2}}+
\epsilon_{\mathrm{int}}+B \\ p_{Q} &=\frac{2 N_{c}}{3} \sum_{i} \int^{k_{F i}} \frac{d^{3} k}{(2 \pi)^{3}} \frac{k^{2}}{\sqrt{k^{2}+m_{i}^{2}}}+p_{\mathrm{int}}-B \\ \mu_{i} &=\sqrt{k_{F i}^{2}+m_{i}^{2}}+\mu_{\mathrm{int}, i} 
\end{align}
The upper limit of the integrals $k_{F i}$ is the Fermi wave number for each species $i$, which, at zero temperature, appropriately terminates the integration over $k$. The first terms in $\epsilon_{Q}$ and in $p_{Q}$ are free Fermi gas (FG) contributions, $\epsilon_{\mathrm{FG}}$ and $p_{\mathrm{FG}}$ respectively, the second terms are due to 
$\mathcal{L}_{\text {int }}$ and $B$ is the bag constant that accounts for the cost of confining the quarks inside a bag. The $m_{i}$ are quark masses, generally taken to be current quark masses. The $u$ and $d$ quark masses are commonly set to zero (because at high density, $k_{F i}$ in these cases far exceed $m_{i}$), whereas that of the $s$ quark is taken at its Particle Data Group (PDG) value. The QCD perturbative calculations of $\epsilon_{\text {pert }}$ and $p_{\text {pert }}$, and the ensuing results for the structure of NSs containing quarks within the cores as well as self-bound strange quark stars are discussed in~\cite{2020NatPh..16..907A}. At leading order of QCD corrections, the results are qualitatively similar to what one obtains by simply using the FG results with an appropriately chosen value of $B$. As results of perturbative calculations are deemed to be valid only for $n_B \geq 40 n_s$, they are dropped in the vMIT model. 
The Lagrangian density from vector  interactions 
\begin{equation}
    \quad \mathcal{L}_{\text {vec }}=-G_{v} \sum_{i} \bar{\psi} \gamma_{\mu} V^{\mu} \psi+\left(m_{V}^{2} / 2\right) V_{\mu} V^{\mu} \,,
\end{equation}
where interactions among the quarks occur via the exchange of a vector-isoscalar meson $V^{\mu}$ of mass $m_{V},$ is chosen in Ref. [54]. 
Explicitly,
\begin{align}
\epsilon_{Q} &=\sum_{i} \epsilon_{\mathrm{FG}, \mathrm{i}}+\frac{1}{2}\left(\frac{G_{v}}{m_{V}}\right)^{2} n_{Q}^{2}+B \\
p_{Q} &=\sum_{i} p_{\mathrm{FG}, \mathrm{i}}+\frac{1}{2}\left(\frac{G_{v}}{m_{V}}\right)^{2} n_{Q}^{2}-B \\
\mu_{i} &=\sqrt{k_{F i}^{2}+m_{i}^{2}}+\left(\frac{G_{v}}{m_{V}}\right)^{2} n_{Q} \,,
\label{vMITeqs}
\end{align}
where $n_{Q}=\sum_{i} n_{i},$ and the bag constant $B$ is chosen appropriately to enable a transition to matter containing quarks. Note that terms associated with the vector interaction above are similar to those in hadronic models. We studied model parameters in a wide range $B^{1 / 4}=(155-180)~\mathrm{MeV}$ and $a=\left(G_{v} / m_{V}\right)^{2}=(0.1-0.3)~\mathrm{fm}^{2}$ and report results for specific values within this range.

\section{Sound Speeds in the Pure and Mixed Phases}
\label{sec4}
As the difference of the adiabatic and equilibrium sound speeds drives the restoring force for $g$-modes, it is instructive to collect some general expressions for these two sound speeds in the pure and mixed phases. For the pure phase of $npe$ matter, these expressions are derived and applied in~\cite{Lai94}, but given their central role in this work, and the fact that we also extend the application to $npe\mu$ and quark matter, we 
detail their derivation below for completeness. For the mixed phase, we derive expressions that have not, to our knowledge, been previously reported in the literature. First, a point of notation: the equilibrium squared sound speed is commonly defined in the literature~\cite{Lai94,Tews_2018,2019PhRvD.100j3022H} by the symbol $c_s^2$, which we reserve here for the squared adiabatic sound speed, as in~\cite{Kantor14}. The equilibrium sound speed is defined by  
\beq
c_e^2 = \frac {dp}{d\epsilon} \,, 
\label{css-pure}
\eeq
where $p$ and $\epsilon$ are the total pressure and energy density in matter
\beq
\epsilon = \epsilon_B\,+\sum\limits_{l=e^-,\,\mu^-} \epsilon_l\,,\quad p = p_B\,+\sum\limits_{l=e^-,\,\mu^-} p_l\,. 
\label{eandpfull}
\eeq
In Eq.(\ref{eandpfull}), the leptonic energies are the $T$=0 degenerate Fermi gas expressions for massive leptons. Being a total derivative, 
the derivative is taken along the curve satisfying both mechanical and chemical equilibrium, i.e., $\beta$-equilibrium conditions hold. In NSM, when only $npe$ are present in equilibrium, the composition at fixed baryon density ($n_n+n_p$) is completely fixed once the proton fraction $x_p$ (=$x_e$ by charge neutrality) is determined. In this case, the squared adiabatic sound speed is defined as 
\beq
c_s^2 = \left( \frac {\partial p}{\partial \epsilon} \right)_x \,, 
\label{css-mixed}
\eeq
where $x=x_p=x_e$. 
In the partial derivative,
the composition is held fixed, i.e., $\beta$-equilibrium conditions are imposed only after all derivatives have been evaluated. The resulting 
distinction between these two speeds plays an important role in  determining the oscillation frequencies of non-radial oscillations such as $g$-modes: 
\beq
\omega^2 \propto \left( \frac {1}{c_e^2} - \frac {1}{c_s^2}  \right)  = \frac {(c_s^2 - c_e^2)}{c_e^2 c_s^2} \,.
\eeq
Note that both the above speeds are dependent on density which varies over a large range in NSM. Furthermore, an individual knowledge of both speeds is required. 
In what follows, we
apply Eqs. (\ref{css-pure}) and (\ref{css-mixed}) to the case of a pure and mixed phase.

\subsection{\label{subsec1}The Pure Phase}
\subsubsection{\label{subsubsec1}
{\bf Sound speeds in $npe$ matter}
}
It is useful to recast the general expressions Eqs.~(\ref{css-pure}) and (\ref{css-mixed}) in terms of derivatives of the individual chemical potentials with respect to density, since such expressions are amenable to both analytical and numerical checks. Without loss of generality, we have 
\beq
c_e^2 = \frac {dp}{d\epsilon} = \left( \frac {dp}{dn_B} \right) \bigg/    \left( \frac {d\epsilon}{dn_B} \right)    \,, 
\eeq
Considering $npe$ matter as an example, differentiating the total energy energy density inclusive of electrons
\beq
\epsilon (n_B,  x) = n_B [ M_B + E (n_B, x) ] 
 \eeq
with respect to $n_B$, we have 
\beq
\left( \frac {d\epsilon}{dn_B} \right)  = \frac {\epsilon}{n_B} + n_B  \left( \frac {dE}{dn_B} \right) \,, 
\label{ceq-deriv}
\eeq
where $E(n_B,x)$ is the energy per baryon. The second term on the 
right hand side of Eq. (\ref{ceq-deriv}) becomes
\beq
\left( \frac {dE}{dn_B} \right) = \left( \frac {\partial E}{\partial n_B} \right)
+ \left( \frac {\partial E}{\partial x} \right)_{n_B} \left( \frac {dx}{dn_B} \right) \,.
\eeq
For the equilibrium sound speed, 
the $\beta$-equilibrium condition $\left( \frac {\partial E}{\partial x} \right)_{n_B} = 0$ yields
\beq
\left( \frac {dE}{dn_B} \right) = \left( \frac {\partial E}{\partial n_B} \right) \,.
\eeq
Thus,
\beq
\left( \frac {d\epsilon}{dn_B} \right)  = \frac {\epsilon}{n_B} + \frac {1}{n_B} \left[ n_B^2  \left( \frac {dE}{dn_B} \right) \right]   
=  \frac {(\epsilon + p)}{n_B} \,.
\label{eplusp}
\eeq
From the thermodynamic identity, using charge neutrality ($x=x_e$) and beta-equilibrium,
\beq
\label{TI1}
\epsilon + p &=& \mu_n n_n + \mu_p n_p + \mu_e n_e \nonumber \\
&=& \mu_n n_B - (\mu_n - \mu_p - \mu_e) n_B =  \mu_n n_B \,,
\eeq
leading to the simple result
\beq
\left( \frac {d\epsilon}{dn_B} \right) = \left( \frac {\partial \epsilon}{\partial n_B} \right)_x = \mu_n 
\label{equals}
\eeq
This implies that
\beq
c_e^2 &\equiv& \frac {dp}{d\epsilon} = \frac {1}{\mu_n}   \left( \frac {dp}{dn_B} \right) =    \frac {1}{\mu_n}  n_B  \left( \frac {d\mu_n}{dn_B} \right)  \, \nonumber \\
&=& \left( \frac {d \ln \mu_n}{d \ln n_B} \right)  \,, 
\eeq
where we have again taken advantage of the thermodynamic identity to relate the required derivative of $p$ to that of $\mu_n$. 

The adiabatic squared sound speed can be expressed as
\beq
c_s^2 &=& \left( \frac {\partial p}{\partial \epsilon} \right)_x  = 
 \left( \frac {\partial p}{\partial n_B} \right)_x \bigg/    \left( \frac {\partial \epsilon}{\partial n_B} \right)_x    \,  \\
 &=& \frac {n_B}{\epsilon+p}  \left( \frac {\partial p}{\partial n_B} \right)_x 
 = \frac {1}{\mu_{\rm avg}} 
 \left( \frac {\partial p}{\partial n_B} \right)_x \,,
 \label{cspd}
\eeq
where we have used the equality in Eq.~(\ref{eplusp}), which is valid at constant composition even in the absence of $\beta$-equilibrium, and introduced an average chemical potential $\mu_{\rm avg}=(\sum_i\mu_in_i)/n_B=(\epsilon+p)/n_B$. Since $p=p_B+p_e$,
\be
c_s^2 = \frac {1}{(\epsilon+p)} 
 \left[\left(u\frac {\partial p_B}{\partial u} \right)_x 
 +\left(u\frac {\partial p_e}{\partial u} \right)_x\right]\,.
 \label{cs2du}
\ee
The required derivatives are analytic:
\beq
\left(u\frac {\partial p_B}{\partial u} \right)_x &=& 
        \frac{2}{9\pi^2}\frac{k_F^5}{E_F} + n_s u \Big[2a_0u + b_0\gamma(\gamma+1)u^{\gamma}\Big]  \nonumber \\
      &+& n_su(1-2x)^2\left\{\frac{2k_F^2}{27E_F}\left(1-\frac{9k_F^2}{10E_F^2}+\frac{3k_F^4}{10E_F^4}\right)\right.  \nonumber \\
      &+& \left. \bigg[2u(a_1-a_0)+b_1\gamma_1(\gamma_1+1)u^{\gamma_1} \right. \nonumber \\
      &-& \left. b_0\gamma(\gamma+1)u^{\gamma}\bigg]\right\} \label{udpbdu} \\
\left(u\frac {\partial p_e}{\partial u} \right)_x &=& \frac{1}{3}n_e\mu_e \label{udpedu} \,.
\eeq

Thus, the difference of squared sound speeds becomes 
\beq
c_s^2 - c_e^2 = \frac {1}{\mu_{\rm avg}}
 \left( \frac {\partial p}{\partial n_B} \right)_x  
 - \frac {1}{\mu_n} \left( \frac {dp}{dn_B}  \right)  \,.
 \label{c2diff1}
\eeq
At this point all the necessary ingredients for the calculation of the speed-of-sound difference are present. It is instructive, however, to obtain a complementary expression in which its physical causes, namely $\beta$-equilibrium and compositional gradients, are made explicit. To that end, we proceed as follows:
Noting that
\beq 
\frac {dp}{dn_B} = \left( \frac {\partial p}{\partial n_B}  \right)_x  +   \left( \frac {\partial p}{\partial x}  \right)_{n_B} \frac {dx}{dn_B} \label{dpdnbfull} \,,
\eeq 
Eq.~(\ref{c2diff1}) can be recast as  
\beq
c_s^2 - c_e^2 &=& 
\left (\frac {1}{\mu_{\rm avg}}  - \frac {1}{\mu_n} \right) 
   \left( \frac {\partial p}{\partial n_B} \right)_x  
   - \frac {1}{\mu_n} \left( \frac {\partial p}{\partial x}  \right)_{n_B} \frac {dx}{dn_B}  \nonumber \\ 
 &=&   -~ \frac {x \tilde\mu}{\mu_{\rm avg}\mu_n} 
 \left( \frac {\partial p}{\partial n_B} \right)_x  
   - \frac {1}{\mu_n} \left( \frac {\partial p}{\partial x}  \right)_{n_B} \frac {dx}{dn_B}    \,
 \label{c2diff2}
\eeq
where $\tilde\mu = \mu_e + \mu_p - \mu_n$. 
Anticipating that $\beta$-equilibrium will be imposed at the end,
we note that the first term above vanishes as 
$\tilde\mu = 0$, which leads to
\beq
c_s^2 - c_e^2 = - \frac {1}{\mu_{n}} \left( \frac {\partial p}{\partial x}  \right)_{n_B} \frac {dx}{dn_B} \,.
\eeq
Utilizing $p = n_B^2 \frac {\partial E}{\partial n_B}$ and interchanging the order of derivatives 
\beq
c_s^2 - c_e^2 = - \frac {1}{\mu_{n}} n_B^2 \frac {\partial}{\partial n_B} \left( \frac {\partial E}{\partial x}  \right)_{n_B} \frac {dx}{dn_B} \,,
\eeq
which can be further rewritten as 
{\footnote{Observe the interesting relation $\frac{\partial p}{\partial x} = n_B^2 \frac{\partial \tilde \mu}{\partial n_B}$, noted also in the context of bulk viscosity studies~\cite{HLY00}.}}
\beq
c_s^2 - c_e^2 = - \frac {1}{\mu_{n}} n_B^2 \left( \frac {\partial \tilde \mu} {\partial n_B} \right)_x \frac {dx}{dn_B} \,.
\label{dc21}
\eeq
It remains now to determine $\frac {dx}{dn_B}$. As
\beq
d\tilde \mu = \left( \frac {\partial \tilde \mu}{\partial n_B} \right)_x dn_B +   \left( \frac {\partial\tilde \mu}{\partial x} \right)_{n_B} dx =0 \,,
\eeq
\beq
\frac {dx}{dn_B} = -  \left( \frac {\partial \tilde\mu}{\partial n_B} \right)_x \bigg/ \left( \frac {\partial \tilde \mu}{\partial x} \right)_{n_B} \,.
\eeq
With this relation,  Eq.~(\ref{dc21}) becomes
\beq
c_s^2 &=&  
c_e^2 + \frac { \left[n_B \left( \frac {\partial \tilde \mu} {\partial n_B} \right)_x \right]^2} { \mu_{n}\left( \frac {\partial \tilde \mu}{\partial x} \right)_{n_B} } \,,
\label{dc22}
\eeq
which illustrates the influence of the density and compositional gradients of the two sound speeds. Thus far, we have simply retraced the steps originally given in~\cite{Lai94} (Sec. 4.2 of this reference).
In $npe$ matter under the constraint of charge neutrality, the independent variables chosen are $n_B$ and $x$, and thus a partial derivative of $\tilde{\mu}$ with respect to $n_B$ ($x$) implies that $x$ ($n_B$) is fixed.  

Casting the expressions for the sound speeds in terms of the chemical potentials is expedient, as illustrated below for the case of $npe$ matter. Note that the average chemical potential $\mu_{\rm avg}=\mu_n$ only in $\beta$-equilibrium. At fixed $x$, with the relation $\tilde \mu = \mu_e - \hat \mu$, 
\beq
n_B \frac {\partial \tilde \mu}{\partial n_B} &=& u \frac {\partial (\mu_e - \hat \mu)} {\partial u} \,.
\label{muhat}
\eeq
For the term in Eq. (\ref{muhat})
 involving electrons, we have
\beq
\mu_e = \hbar c ~(3\pi^2n_s u)^{1/3} x^{1/3} \quad {\rm and} \quad  
u \frac {\partial \mu_e}{\partial u} =  \frac {\mu_e}{3} \,,
\eeq
while for baryons,{\footnote{Details of the derivatives of the kinetic part of the symmetry energy are given in Appendix A.}}  
\beq
\hat \mu &=& \mu_n - \mu_p = 4 S_2(u) (1-2x) \quad {\rm with} \nonumber \\
S_2(u) &=&  S_{2k} + S_{2i} = \frac {k_F^2}{6E_F} \nonumber \\ 
&+& (a_1 - a_0) u + b_1 u^{\gamma_1} - b_0 u^\gamma \,, \nonumber \\
u S_{2k}^\prime  &=& \frac 13 \cdot 2 S_{2k} \left[ 1 -  18 \left( \frac {S_{2k}}{k_F}  \right)^2  \right], \nonumber \\
{\rm and} \quad 
u S_{2i}^\prime &=& (a_1 - a_0) u + \gamma_1 b_1 u^{\gamma_1} - \gamma b_0 u^\gamma \,, \nonumber \\
u S_{2}^\prime &=& u S_{2k}^\prime + u S_{2i}^\prime.
\eeq
Putting together the above results, we have 
\beq
n_B \frac {\partial \tilde \mu}{\partial n_B} = \frac {\mu_e}{3} - 4 (1-2x)~ u S_{2}^\prime \,.
\label{nbdnbdu}
\eeq
Derivatives with respect to $x$ of $\tilde \mu $ at fixed density are also straightforward. For 
\beq
\frac {\partial \tilde \mu}{\partial x} =  \frac {\partial (\mu_e - \hat \mu)}{\partial x} \,,
\eeq
we note that 
\beq
\frac {\partial \mu_e}{\partial x} &=& \frac 13 \frac {\mu_e}{x} \quad {\rm and} \quad  \frac {\partial \hat \mu}{\partial x} = -8S_2(u) \quad {\rm so~ that} \nonumber \\  
\frac {\partial \tilde \mu}{\partial x} &=& \frac 13 \frac {\mu_e}{x} + 8 S_2(u) \,. 
\label{dmudx}
\eeq
The equivalence of Eqs. (\ref{cs2du}) and (\ref{dc22}) is established analytically in Appendix A.5. 

\subsubsection{Sound speeds in $npe\mu$ matter}

Going beyond the results in~\cite{Lai94}, one way to include muons is by choosing $n_B$, $x=x_e+x_\mu$ and $x_\mu\equiv y$ as the independent variables. The formal expression for the squared adiabatic sound speed remains the same as in $npe$ matter, i.e., Eq.~(\ref{cs2du}) but now $\left(u~\partial p_e/\partial u\right)_x$ [Eq.~(\ref{udpedu})] is replaced by
\be
\left(u\frac {\partial p_{lep}}{\partial u} \right)_{x,x_{\mu}} = 
        \frac{1}{3}n_e\mu_e + \frac{1}{3}n_{\mu}\left(\frac{\mu_{\mu}^2-m_{\mu}^2}{\mu_{\mu}}\right) \,.
        \label{udlepdu}
\ee
where $lep = e^-, \mu^-$.

Furthermore, by retracing the steps leading to Eq.~(\ref{dc22}), its $npe\mu$ equivalent is obtained as
\be
c_s^2 - c_e^2  =  -\frac{1}{\mu_n}\left(\left.\frac{\partial P}{\partial x}\right|_{n_B,y}\frac{dx}{dn_B}
                  +\left.\frac{\partial P}{\partial y}\right|_{n_B,x}\frac{dy}{dn_B}\right) 
                  \label{csmince}
\ee      
with~\footnote{The intermediate steps leading to Eqs. (\ref{dxdnmu})-(\ref{dydnmu}) are detailed in Appendix. A.6 }
\beq
\frac{dx}{dn_B} &=& \frac{\left.\frac{\partial \tilde\mu_x}{\partial y}\right|_{n_B,x}
                        \left.\frac{\partial \tilde\mu_y}{\partial n_B}\right|_{x,y}
                       -\left.\frac{\partial \tilde\mu_y}{\partial y}\right|_{n_B,x}
                       \left.\frac{\partial \tilde\mu_x}{\partial n_B}\right|_{x,y}}
                       {\left.\frac{\partial \tilde\mu_x}{\partial y}\right|_{n_B,x}
                       \left.\frac{\partial \tilde\mu_y}{\partial x}\right|_{n_B,y}
                       -\left.\frac{\partial \tilde\mu_y}{\partial y}\right|_{n_B,x}
                       \left.\frac{\partial \tilde\mu_x}{\partial x}\right|_{n_B,y}}  \label{dxdnmu}\\
\frac{dy}{dn_B} &=& \frac{\left.\frac{\partial \tilde\mu_x}{\partial x}\right|_{n_B,y}
                         \left.\frac{\partial \tilde\mu_y}{\partial n_B}\right|_{x,y}
                       -\left.\frac{\partial \tilde\mu_y}{\partial x}\right|_{n_B,y}
                       \left.\frac{\partial \tilde\mu_x}{\partial n_B}\right|_{x,y}}
                       {\left.\frac{\partial \tilde\mu_x}{\partial x}\right|_{n_B,y}
                       \left.\frac{\partial \tilde\mu_y}{\partial y}\right|_{n_B,x}
                       -\left.\frac{\partial \tilde\mu_y}{\partial x}\right|_{n_B,y}
                       \left.\frac{\partial \tilde\mu_x}{\partial y}\right|_{n_B,x}} \label{dydnmu} ~.
\eeq       
The chemical potentials $\tilde\mu_x = \mu_p+\mu_e-\mu_n$ and $\tilde\mu_y = \mu_{\mu}-\mu_e$ are zero in $\beta$-equilibrated matter. Equations~(\ref{csmince})-(\ref{dydnmu}), while demonstrating that compositional gradients are at the core of g-mode oscillations, are 
lengthy
and computationally 
more involved
compared to the 
direct calculation of the adiabatic sound speed in $npe\mu$ matter using Eqs.~(\ref{udpbdu}) and (\ref{udlepdu}). For the sake of completeness, we provide here the explicit expressions for the adiabatic sound speed in $npe\mu$ matter arising from ~(\ref{csmince})-(\ref{dydnmu}) which are in excellent numerical agreement with the more direct method:

\beq
c_s^2=c_e^2+\frac{1}{\mu_n}\Big(T_1+T_2+T_3+T_4\Big)
\eeq
where $T_j=N_j/D$ with 
\begin{align}
N_{1}&=\left[\frac{\mu_{e}}{3}-4(1-2x)~u S_{2}^{\prime}\right]^{2}   \\ \nonumber 
N_{2}&=\left[\frac{\mu_{e}}{3}-4(1-2x)~u S_{2}^{\prime}\right] 8S_2x_e\left[\frac{k_{F_{\mu}}}{k_{F_e}}-\frac{x_{\mu}}{x_e}\right] \\ \nonumber 
N_{3}&=\left[\frac{k_{F_{\mu}}^2}{3\mu_e}-4(1-2x)~u S_{2}^{\prime}\right]\left(\mu_{e}+8S_2x_e\right) \left[\frac{x_{\mu}}{x_e}-\frac{k_{F_{\mu}}}{k_{F_{e}}}\right] \\ \nonumber
N_{4}&=\left[\frac{k_{F_{\mu}}^2}{3\mu_e}-4(1-2x)~u S_{2}^{\prime}\right]\frac{k_{F_{\mu}}}{k_{F_{e}}}\left[\frac{\mu_{e}}{3}-4(1-2x)~u S_{2}^{\prime}\right] \\ \nonumber 
D&=\left[\frac{\mu_{e}}{3 x_{e}}+8 S_{2}\left(1+\frac{k_{F_\mu}}{k_{F e}}\right)\right] 
\end{align}
and $k_{F_e}=\mu_e$ (massless electrons) and $k_{F_{\mu}}$=$\sqrt{{\mu_\mu}^2-m_{\mu}^2}$.
These equations explicitly display the connection to the nuclear symmetry energy $S_2$ and its density derivative $S_2^\prime$. \\

At the muon threshold ($k_{F_{\mu}}, x_{\mu}$=0 $\Rightarrow$ $x$=$x
_e$), it is easy to see that $N_2,N_3,N_4$=0, while $N_1(\equiv N_1^{npe}$) recovers Eq.~(\ref{dc22}) for $npe$ matter. At extremely high baryon density, muons are ultra-relativistic ($\mu_{\mu}$=$k_{F_{\mu}}$=$k_{F_e}$, $x_{\mu}$=$x_e$=$x/2$) so that $N_2,N_3$=0, $N_1$=$N_4$=$N_1^{npe}/2$ and the total leptonic contribution to the sound speed is equally divided between electrons and muons. 
\\
\subsubsection{\label{subsubsec2} {\bf Sound speeds in} Quark {\bf Matter}}


We now move to a discussion of sound speeds in dense quark matter at zero temperature. For the pure quark phase, 
the difference of the two sound speeds has been computed to leading order in the quark mass~\cite{Olinto96,Wei_2020} using the non-interacting 3-flavor FG model with massive quarks (see Sec.\ref{vMIT-EOS}). These expressions reveal that for the non-interacting FG model, a non-zero quark mass is necessary to support $g$-modes. This is because a system of massless $uds$ quarks is charge neutral with equal numbers of each flavor at any density; effectively, there is no change in composition with density to drive the $g$-mode.



To leading order in the $s$-quark's mass $m_s$, the \bv\,frequency is~\cite{Wei_2020}
\beq
N_q \simeq\left(\frac{g}{2 \pi c_{e}}\right)\left(\frac{m_{s}^{2}} {\sqrt{B}}\right) \,,
\eeq
where $c_e^2=dp_q/d\epsilon_q$ is the equilibrium squared sound speed in QM\footnote{Numerically, $N_q\approx 100$ Hz for a current quark mass $m_s\approx 100$ MeV, but the effect of interactions in addition to this yields significantly lower values for $N_q$~\cite{Wei_2020}.}. 
It is possible to obtain an exact expression for $c_e^2$ and    $c_s^2$ 
in QM 
for the FG model, and also for the vMIT model, as we show below. 

The equilibrium sound speed may be simply calculated by numerically evaluating $c_{e,vMIT}^2 = dp/d\epsilon$ in the pure quark phase. However, additional insight into its compositional structure is gained by expressing it in terms of the various chemical potentials involved. Starting from the relation (valid in $\beta$-equilibrium)
\beq
\mu_n = 2 \mu_d + \mu_u = (2 \mu_d^{\ast} + \mu_u^{\ast}) + 3 a n_Q\,,
\eeq
where $\mu_f^{\ast}=\sqrt{k_{F_f}^2+m_f^2}$ for a quark of flavor $f$, and using $c_e^2=d \ln \mu_n / \ln n_B$,\footnote{In charge neutral and $\beta$-equilibrated matter, 
$\mu_B= \sum_f x_f \mu_f + \sum\limits_
{\ell=e,\mu} x_\ell \mu_\ell = \mu_n$ as in the nucleonic phase.}  we obtain
\beq
c_{e,vMIT}^2 =  \frac{1}{\mu_n} &\bigg[&
\frac{1}{3} 
\left\{2\mu_d^{\ast}\left(1-\frac{m_d^2}{\mu_d^{\ast^2}}\right)
\frac {d\ln n_d}{d\ln n_B} \right. \nonumber \\
&+&  
\left.\mu_u^{\ast}\left(1-\frac{m_u^2}{\mu_u^{\ast^2}}\right)
\frac {d\ln n_u}{d\ln n_B} \right\} \nonumber \\ 
&+& 3a n_Q\bigg] \,.
\label{ce2vMIT}
\eeq
Contributions from the leptons are implicitly included in the above expression.

For the non-interacting FG model, the pressure $p=\sum_f p_{\rm FG}(\mu_f,\mu_e)$.   
Introducing the partial fractions $x_f=n_f/n_B$, where $n_f=(\mu_f^2-m_f^2)^{3/2}/\pi^2$ and $n_e=\mu_e^3/3\pi^2$, the partial derivative of pressure with respect to baryon density in the definition of the adiabatic sound speed in Eq.~(\ref{cspd}) can be re-expressed in terms of partial derivatives with respect to the various chemical potentials, yielding
\beq
\label{cmixed-MIT}
c_{s,{\rm FG}}^2=\frac{1}{\mu_{\rm avg}}\left[\sum_f\frac{1}{3}\mu_fx_f\left(1-\frac{m_f^2}{\mu_f^2}\right)+\frac{1}{3}\mu_ex_e\right] \,,
\eeq
where $\mu_{\rm avg}=(\sum\limits_
{f=u,d,s,e}n_f\mu_f)/n_B$.  
Note that if all $m_f$ = 0 (i.e, one is in a charge neutral phase with $x_e$ = 0), $c_{s,{\rm FG}}^2=c_{e,{\rm FG}}^2=1/3$ and there can be no $g$-modes. Inclusion of ${\cal O}(\alpha_s)$ corrections to this model does not change the fact that a non-zero quark mass is necessary for $g$-modes.

In the vMIT model given by Eqs. (\ref{vMITeqs}), 
$\mu_f^{\ast}=\sqrt{k_{F_f}^2+m_f^2}=\mu_f-an_Q$, and as was done for the FG model, we compute partial derivatives with respect to $\mu_f^{\ast}$, noting that $n_f(\mu_f)$ = $n_{\rm FG}(\mu_f^{\ast})$. The resulting expression for the adiabatic sound speed in the vMIT model is
\beq
c_{s,vMIT}^2=\frac{1}{\mu_{\rm avg}}&\bigg[&\sum_f\frac{1}{3}\mu_f^{\ast}x_f\left(1-\frac{m_f^2}{\mu_f^{\ast^2}}\right) \nonumber \\
&+&\frac{1}{3}\mu_ex_e+3a n_Q\bigg] \,,
\label{cs2vMIT}
\eeq
where all quantities $\mu_f, \mu_e, x_e, x_f$ are equilibrium values\footnote{Inclusion of muons is straightforward and adds a term, $\frac{1}{3}\mu_{\mu}x_{\mu}\left(1-\frac{m_{\mu}^2}{\mu_{\mu}^{\ast^2}}\right)$, on the right hand side of Eq.~(\ref{cmixed-MIT}).}. If we switch off interactions ($a\rightarrow$ 0), we recover results of the non-interacting FG model. Interestingly, if we retain the interaction term, but set all quark masses equal or to zero (implying that $x_e$ = 0), we find that $c_{s,vMIT}^2=c_{e,vMIT}^2$ so stable $g$-modes are not supported in the pure quark phase. Therefore, while both sound speeds are modified by interactions, e.g.,
\beq
c_{s,vMIT}^2=\frac{1}{\mu_{\rm avg}}\left[\mu_q^*+3an_Q\right]\neq c_{s,{\rm FG}}^2 \,,
\eeq
at asymptotically high density where quark masses are negligible, there can be no $g$-modes in quark matter in the vMIT model~\footnote{$g$-modes would still exist in a mixed phase of vMIT quark matter and nucleons as the electron fraction would vary from $\beta$-processes involving nucleons.}

Note that when all chemical potentials and partial fractions are set to their equilibrium values for $c_{s,vMIT}^2$ in the pure phase, $\mu_{\rm avg} = \mu_n$. A comparison of Eqs. (\ref{ce2vMIT}) and (\ref{cs2vMIT}) reveals the differences between the two sound speeds. While effects of interactions enter in the same formal way for the two squared speeds, the occurrence of the logarithmic derivatives of the quark densities distinguishes $c_{e,vMIT}^2$ from 
$c_{s,vMIT}^2$ which features the partial fractions $x_f$.
This difference is the principal reason for the latter to become larger than the former. In both cases, the $d$-quark contributions are larger than those of $u$ and $s$ quarks.

\subsection{\label{subsec2}{Sound Speeds in} the Mixed Phase}


 Once we have expressions for the sound speed in a pure phase of quarks or nucleons, it is possible to compute the sound speed in the mixed phase of the two, obtained from a Gibbs construction. The only information required, other than the sound speeds in the pure phases, is the partial phase fraction of quarks $\chi$ at any density. It is more convenient to begin with the reciprocal relation 
\beq
\label{cemix}
\frac{1}{c_{e,{\rm mix}}^2}=\left(\frac{d \epsilon_{\rm mix}}{dp_{\rm mix}}\right)\,.
\eeq
In a Gibbs mixed phase, the pressures in the two phases are equal, while the energy density is a proportional mix of the quark ($q$) and nucleonic/hadronic ($h$) phases: 
$\epsilon_{\rm mix}=(1-\chi)\epsilon_h+\chi\epsilon_q$. Substituting this in Eq.(\ref{cemix}) gives

\beq
\frac{1}{c_{e,{\rm mix}}^2}=\frac{(1-\chi)}{c_{e,h}^2}+\frac{\chi}{c_{e,q}^2}+(\epsilon_q-\epsilon_h)\frac{d\chi/dn_B}{dP/dn_B}\,.
\label{cemix2}
\eeq
The derivatives in Eq.(\ref{cemix2}) must be computed numerically after solving for $\chi$, hence afford no particular advantage over a direct numerical computation of the sound speed from Eq.(\ref{cemix}) itself. However, note that the last term in Eq.(\ref{cemix2}) is always positive in the mixed phase.

As before, the general definition of the adiabatic sound speed applies to the mixed phase 
\beq
c_{s,{\rm mix}}^2=\left(\frac{d p_{\rm mix}}{d \epsilon_{\rm mix}}\right)_{x_i={\rm const.}=x_{i,{\rm eq.}}}\,,
\eeq
and the thermodynamic identity becomes $\epsilon_{\rm mix}+ p_{\rm mix}$ = $\sum_in_i\mu_i$ = $n_B\mu_{\rm avg}$. Noting that the derivatives
$\partial\epsilon_h/\partial_{n_{B,h}}$ and
$\partial\epsilon_q/\partial_{n_{B,q}}$ are equal to the respective $\mu_{\rm avg}$, it is once again more convenient to begin with the reciprocal relation 
\beq
\frac{1}{c_{s,{\rm mix}}^2}=\left(\frac{d \epsilon_{\rm mix}}{dp_{\rm mix}}\right)_{x_i={\rm const.}=x_{i,{\rm eq.}}}\,,
\eeq
and use the chain rule to compute derivatives with respect to density. This leads to
\beq
\left(\frac{d \epsilon_{\rm mix}}{dp_{\rm mix}}\right)_{x_i=x_{i,{\rm eq.}}}=\frac{(1-\chi)\mu_{\rm avg}}{\left(\frac{\partial p_h}{\partial n_B,h}\right)}+\frac{\chi\mu_{\rm avg}}{\left(\frac{\partial p_q}{\partial n_B,q}\right)}\,,
\eeq
which, using Eq. (\ref{cspd}), becomes
\beq
\frac{1}{c_{s,{\rm mix}}^2}=\frac{(1-\chi)}{c_{s,h}^2}+\frac{\chi}{c_{s,q}^2}\,.
\label{csmix2}
\eeq
Comparing Eqs.~(\ref{cemix2}) and (\ref{csmix2}), and to the extent that the two sound speeds in the pure hadronic/quark phase are almost equal, we expect that the last term in 
Eq.~(\ref{cemix2}) which tracks the rapidly changing composition in the mixed phase, is mainly responsible for $c_{s,\rm mix}^2>c_{e,\rm mix}^2$. The more rapid the appearance of new chemical species and the softer the mixed phase, the larger the \bv\, frequency will be.

Furthermore, as will become evident from our results in Sec.~\ref{sec5}, the adiabatic sound speed is continuous across the transition to and from the mixed phase, while the equilibrium sound speed has a slight jump to accommodate the derivative of $\chi$. The reciprocal relation for the adiabatic sound speeds is reminiscent of the addition of resistors in a parallel circuit, with voltage as pressure and current as energy density. Such impedance analogies arise commonly in electrical engineering when modeling the behavior of transducers.

\section{Results}
\label{sec5}

\subsection{Structural properties, sound speeds and the \bv ~frequency  } 

\begin{figure}[]
\includegraphics[scale=0.24]{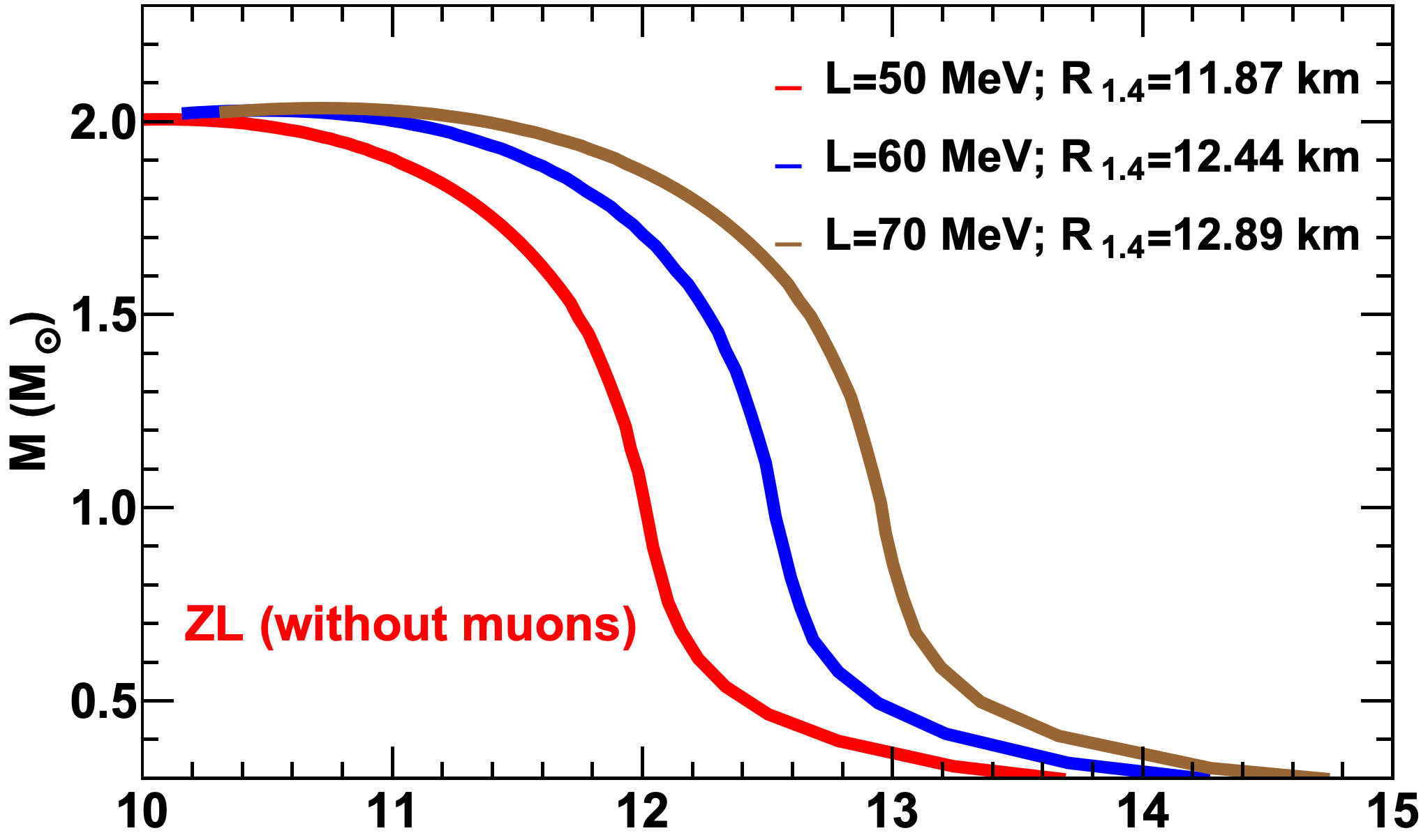}\\[-3.0ex]
\includegraphics[scale=0.24]{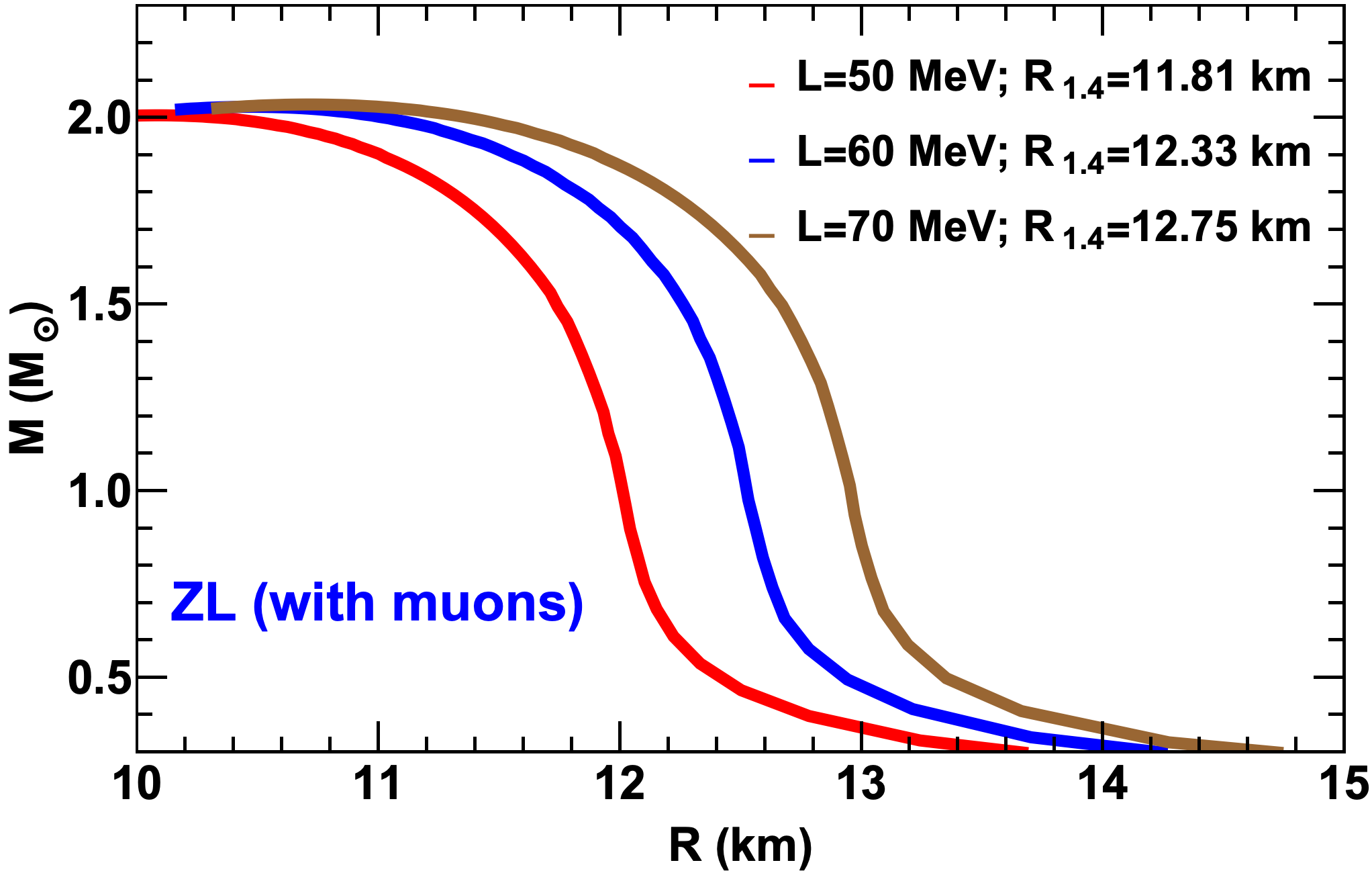}
%
\caption{\label{fig:1} Mass-radius curves for the ZL EOS without and
with  muons. Configurations with muons are slightly more compact, but both cases support $M_{\rm max}\simeq 2M_{\odot}$.
Except for $L$, the EOS parameters are $K_0=220$ MeV, $S_v=31$ MeV, and $\gamma_1=1.6$ for all curves. 
}
\label{fig:MR-ZL-EOS}
\end{figure}

Figure~\ref{fig:MR-ZL-EOS} shows $M$-$R$ curves for ZL EOSs with and without muons for the indicated parameters in the 
caption.
The radii of $\sim 1.4~M_\odot$ stars, $R_{1.4}$, for the different models shown lie within the bounds  inferred from 
available data. For example, data from X-ray observations have yielded $R_{1.4} = 9$-$14$ km for 
canonical masses of $\sim 1.4~M_\odot$~\cite{2012ARNPS..62..485L,2016ARA&A..54..401O,Ozel_2016}. Measured tidal deformations from gravitational wave data in the binary NS merger 
GW170817 give 8.9-13.2 km for binary masses in the range 1.36(1.17)-1.6(1.36) $M_\odot$~\cite{PhysRevLett.121.091102}, whereas for the same masses Capano et al.~\cite{2020NatAs...4..625C} report $11\pm 1$ km. X-ray pulse analysis of NICER data from
PSR J0030+0451 by Miller et al. (2019)~\cite{Miller-nicer} 
finds $13.02^{+1.14}_{-1.19}$ km for $M=1.44\pm 0.15~M_\odot$, whereas for the same star Riley et al. (2019)~\cite{Riley-nicer} obtain $12.71^{+1.14}_{-1.19}$ km and 
$M=1.34^{+0.15}_{-0.16}~M_\odot$.  
The  maximum masses ($\simeq 2M_\odot$) of these EOSs\footnote{By adjusting the constants of the ZL EOS to make the EOS stiffer (yet causal) at supra-nuclear densities, masses larger than $2M_\odot$ can be obtained; an example will be shown later.} are also within the uncertainties of high mass NSs which range from $1.908\pm 0.016~M_\odot$ to $2.27^{+0.17}_{-0.15}~M_\odot$~\cite{2010Natur.467.1081D,2016ApJ...832..167F,Arzoumanian_2018,2020NatAs...4...72C,Linares_2018}.
Although differences in $R_{1.4}$
with and without muons for a given EOS are small, the appearance of muons in the star leads to  distinct features in the \bv ~frequency (see below).

\begin{figure}[hbtp]
\leavevmode
\includegraphics[scale=0.41]{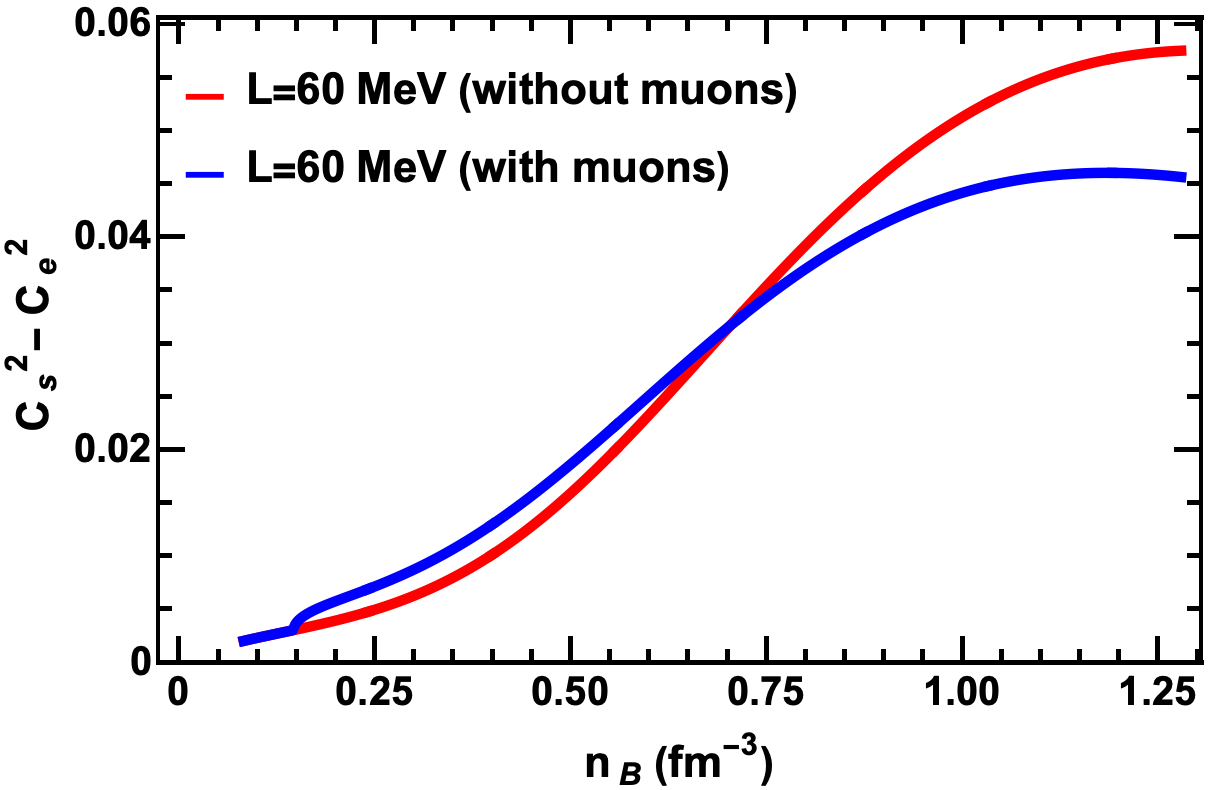}
\caption{\label{fig:2} Difference between the adiabatic and equilibrium squared sound speeds (normalized to the squared speed of light) for the ZL EOS ($K_0$=220 MeV, $S_v=31$ MeV, $L=60$ MeV and $\gamma_1$=1.6) without and with muons.}
\label{fig:c^2diff}
\end{figure}

In Fig.~\ref{fig:c^2diff}, differences in the two squared sound speeds are shown as a function of $n_B$ with and without muons for the ZL EOS with $L=60$ MeV. The small jump at $n_B\simeq 0.14~{\rm fm}^{-3}$, the density at which muons appear, is caused by a sudden drop in the equilibrium sound speed. The differences at large densities are due to the increasing concentration of muons.

\begin{figure}[]
\leavevmode
\includegraphics[scale=0.455]{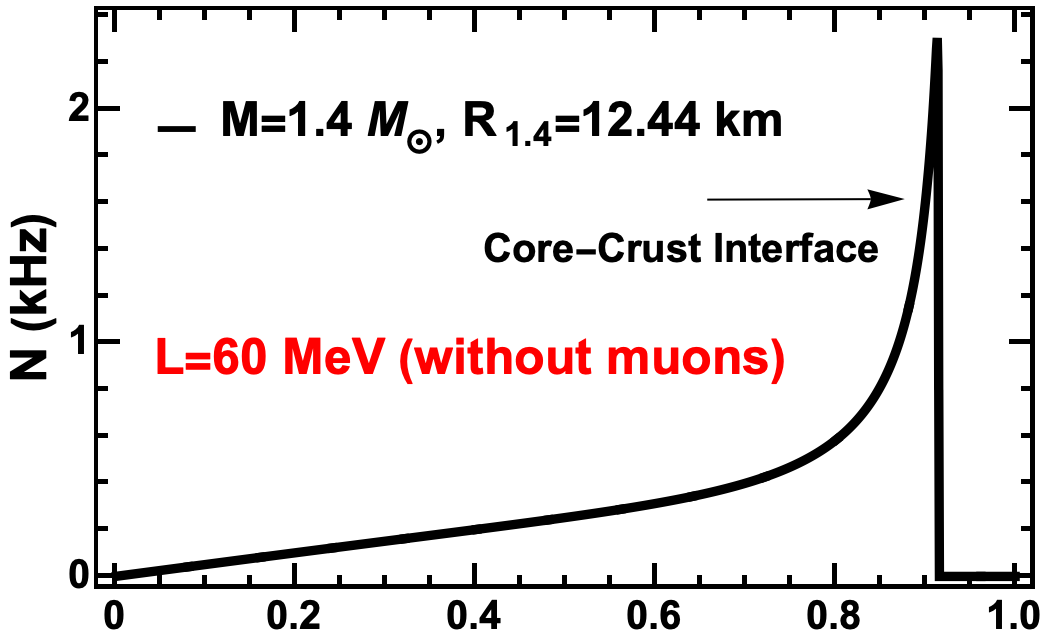}\\
[-3.2ex]
\includegraphics[scale=0.455]{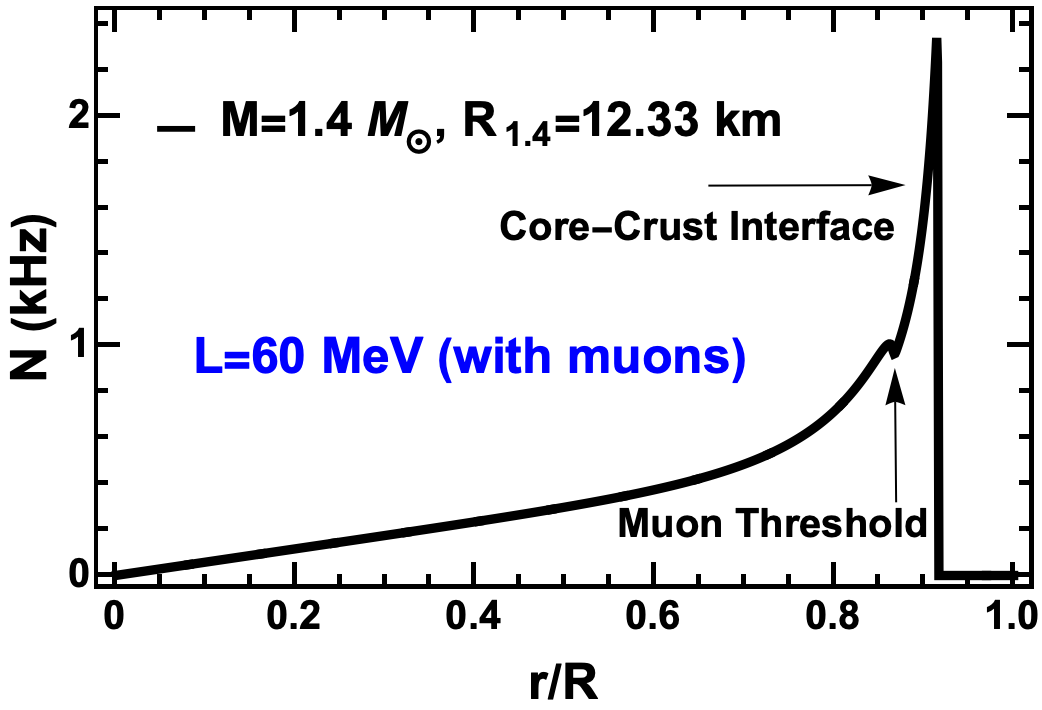}\\
[-2ex]
\caption{\label{fig:3} The \bv\, frequency in the NS for the ZL EOS ($K_0$=220 MeV, $S_v=31$ MeV, $L=60$ MeV, and  $\gamma_1$=1.6) without and with muons.
}
\label{fig:BV-freq}
\end{figure}

\begin{figure}[]
\includegraphics[scale=0.57]{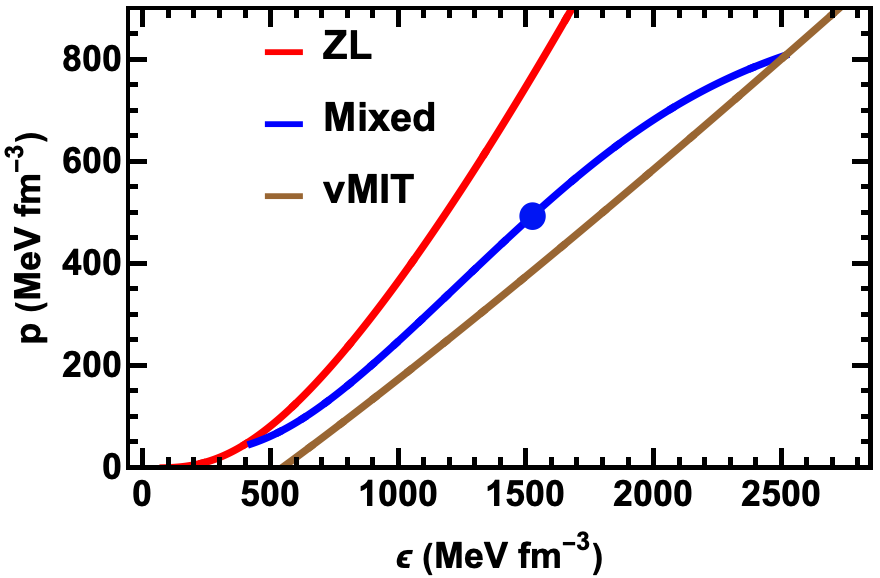}\\[0.5ex]
\caption{\label{fig:5} EOS for the mixed phase of nucleons and quarks (middle curve) using the Gibbs construction. For the ZL EOS without muons, $K_0$=220 MeV, $S_v=31$ MeV, $L=60$ MeV, and $\gamma_1$=1.6. 
Parameters for the vMIT EOS are: ($m_u,m_d,m_s$)=(5,7,150) MeV, 
$B^{1/4}$=180 MeV and $a$=0.1. The circle indicates the central $p$ and $\epsilon$ of the maximum mass star ($n_{c,{\rm max}}/n_s$=7.63, for $M_{\rm max}/M_{\odot}$=1.82).}
\label{fig:EOS-Mixed}
\end{figure}

Figure  \ref{fig:BV-freq} shows the \bv\, frequency 
$N$ vs 
$r/R$ in the star. In the results shown, the crust is assumed to be a homogeneous fluid for simplicity, hence $N$ vanishes there. The location 
where muons begin to appear is signaled by the small kink in the bottom panel. Overall, 
$N$ is slightly larger with muons in the density range in the core of a 1.4$M_{\odot}$ star, consistent with Fig.~\ref{fig:c^2diff}. This has a proportional impact on the $g$-mode frequency as shown in the next section. 

The EOS of the mixed phase following the Gibbs construction, and the ZL EOS for the nucleonic sector and the vMIT EOS for the quark sector 
is shown in Fig.~\ref{fig:EOS-Mixed}. 
The ZL EOS does not include the small effect of muons. In the quark sector, 
muons have not been included since their 
impact relative to quarks is tiny.

\begin{figure}[]
\includegraphics[scale=0.57]{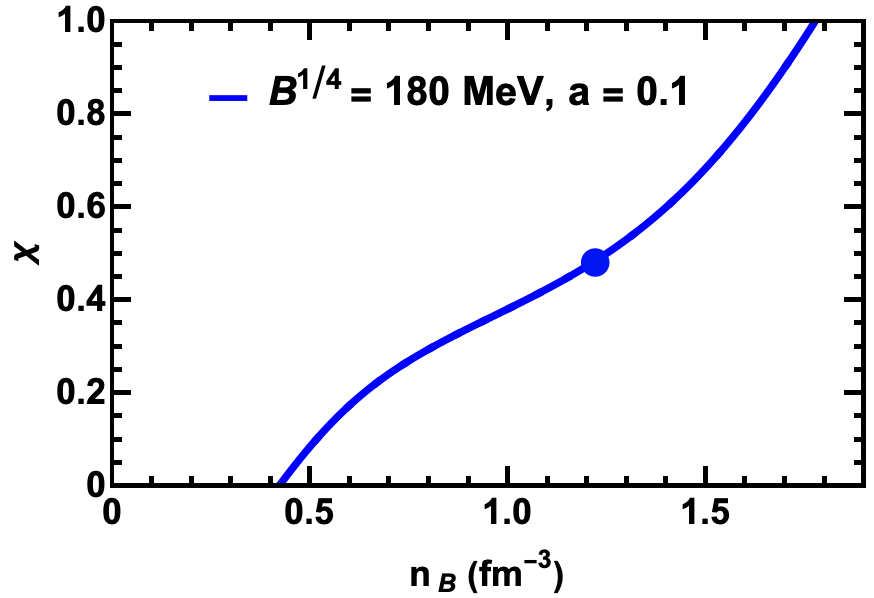}\\[-2ex]
\caption{Quark fraction vs $n_B$ corresponding to Fig.~\ref{fig:EOS-Mixed}. The circle indicates the central density of the maximum mass star ($n_{c,{\rm max}}$=1.22~${\rm fm}^{-3}$ for $M_{\rm max}/M_{\odot}$=1.82).}
\label{fig:Qfrac}
\end{figure}

The compositional change in the mixed phase is indicated by the quark fraction $\chi$ in Fig.~\ref{fig:Qfrac}. The 
steep rise of $\chi$
from
the onset indicates the sort of rapid compositional change that can impact the $g$-mode frequency. A similar effect has been reported~\cite{Dommes15} in the context of the appearance of strange baryons (e.g. hyperons), which is not a phase transition. Note, that for the EOSs considered, the central density of the maximum mass star, indicated by the filled circle on the curve, lies in the mixed phase so that the pure quark phase is not realized.

\begin{figure}[]
\includegraphics[scale=0.53]{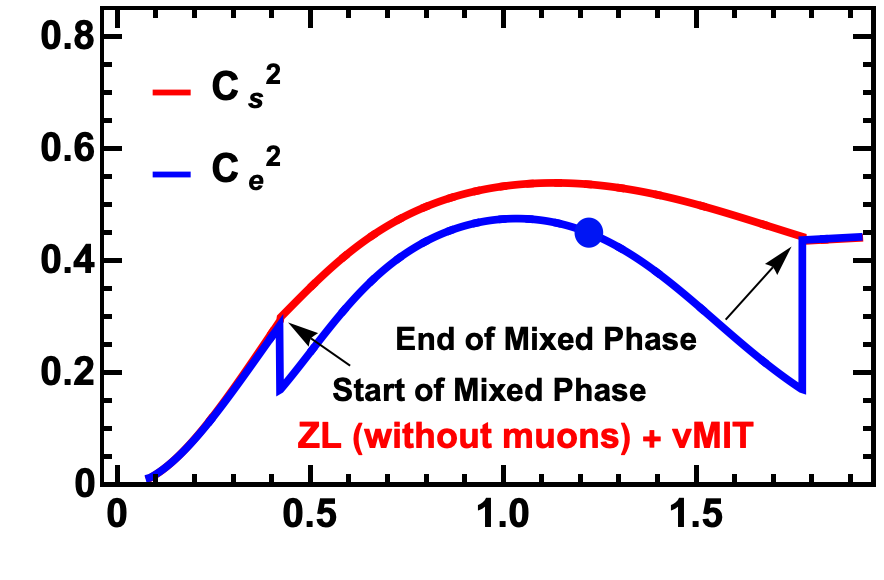}\\[-6.0ex]
\includegraphics[scale=0.51]{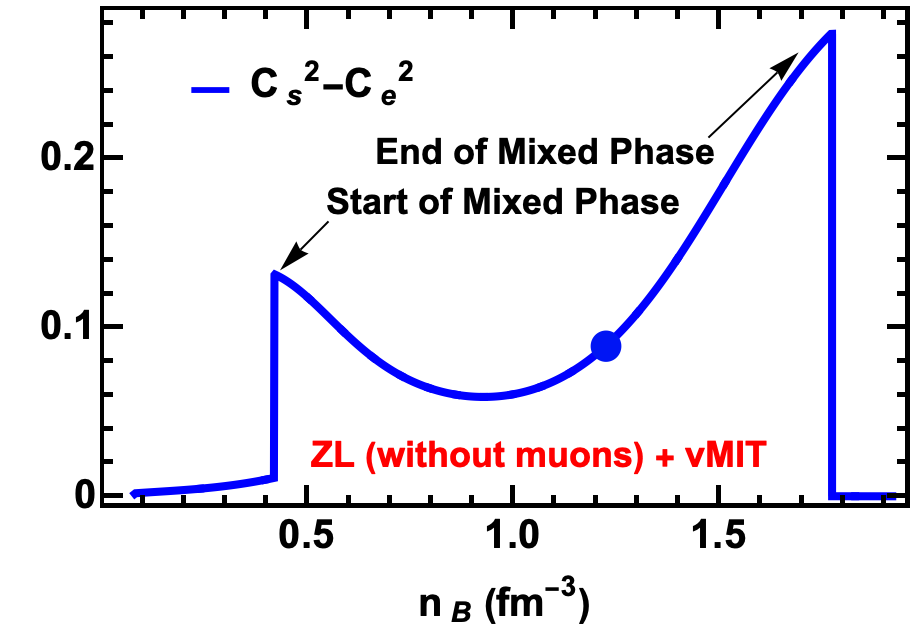}\\[-2ex]
\caption{\label{fig:6}  
The two sound speeds (top panel) and their differences (bottom panel) in the mixed phase for the EOS parameters corresponding to Fig.~\ref{fig:EOS-Mixed}.
The pure quark phase is not achieved prior to the maximum mass 
in this case.
The termination at $n_B$=0.08 fm$^{-3}$ demarcates the core-crust boundary, since we assume $c_s$=$c_e$ in the core. Both sound speeds take much smaller values in the crust than in the core. The circle indicates the central density of the maximum mass star ($n_{c,{\rm max}}$=1.22~${\rm fm}^{-3}$ for $M_{\rm max}/M_{\odot}$=1.82). }
\label{fig:c2diff-Mixed}
\end{figure}

Figure~\ref{fig:c2diff-Mixed} shows results for the individual sound speeds and their differences for the mixed phase.  The two sound speeds in the mixed phase behave very differently. Specifically, the equilibrium sound speed suddenly drops (rises) at the onset (end) of the mixed phase, whereas the adiabatic sound speed varies smoothly. 

\begin{figure}[htbp]
\centering
\includegraphics[scale=0.54]{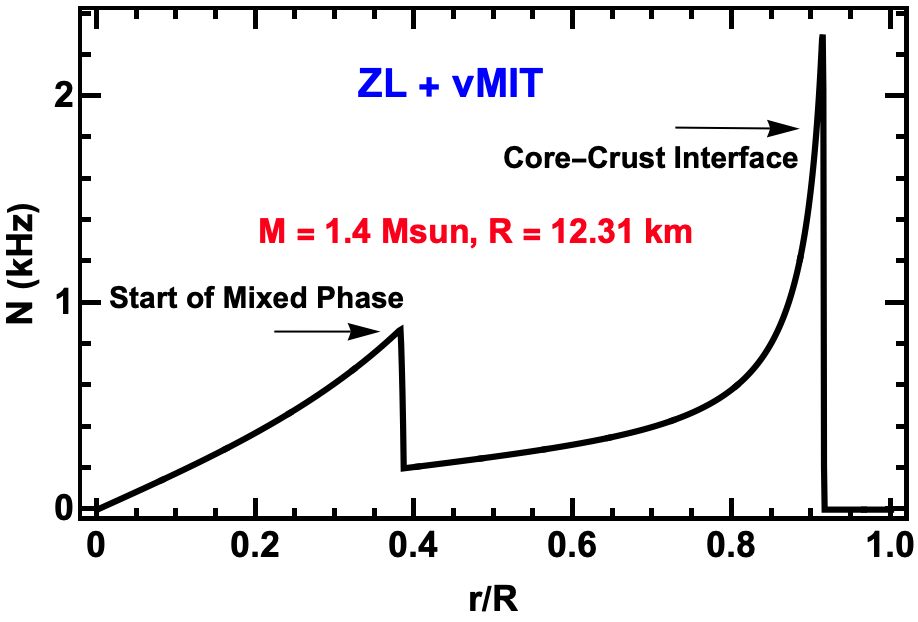}
\caption{The \bv\, frequency in a hybrid star of mass $1.4~M_\odot$. The ZL EOS does not include muons and parameters for the nuclear and quark EOS are as 
in Fig.~\ref{fig:EOS-Mixed}. Quarks enter at
$n_B \simeq$ 0.42 fm$^{-3}$ 
corresponding to $r/R$ = 0.383, and the mixed phase extends beyond the central density. The value of $N$ decreases towards the core due to the decreasing value of $g$, even as the sound speed difference does not change much.}
\label{fig:BV-Mixed}
\end{figure}

The \bv\, frequency of a $1.4~M_\odot$ hybrid star 
is shown in Fig.~\ref{fig:BV-Mixed}.  Note the broader width of the peak when quarks enter, and its location in denser regions of the star, as compared to the nucleonic stars depicted in Fig.~\ref{fig:BV-freq}. This explains why the $g$-mode, which is a long-wavelength global oscillation, is strongly impacted by the mixed phase (see results in the next section).

\begin{figure*}[htbp]
\leavevmode
\parbox{0.48\hsize}{
\includegraphics[width=8.5cm,height=5.5cm]{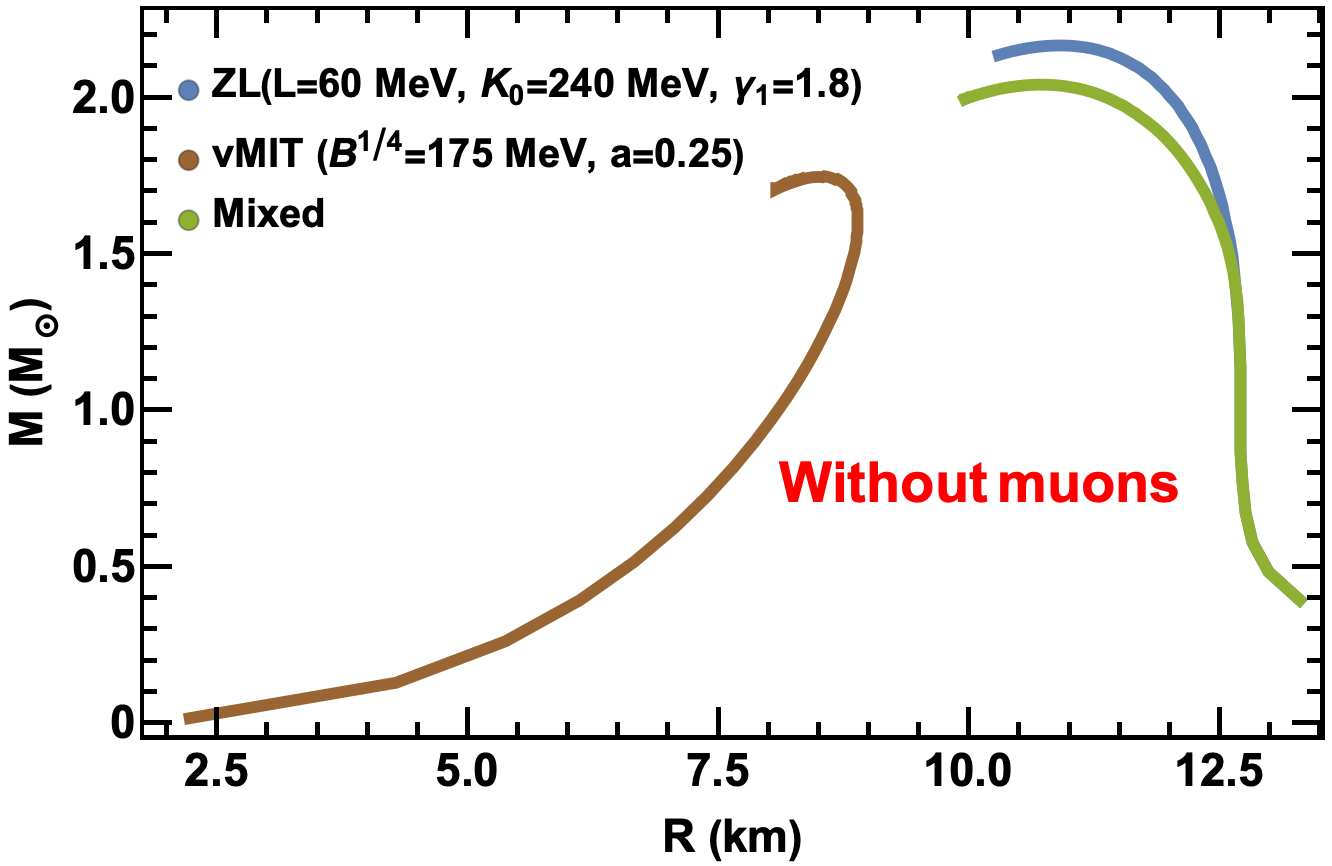}\\[-2ex]
}\parbox{0.52\hsize}{
\includegraphics[width=8.7cm,height=5.5cm]{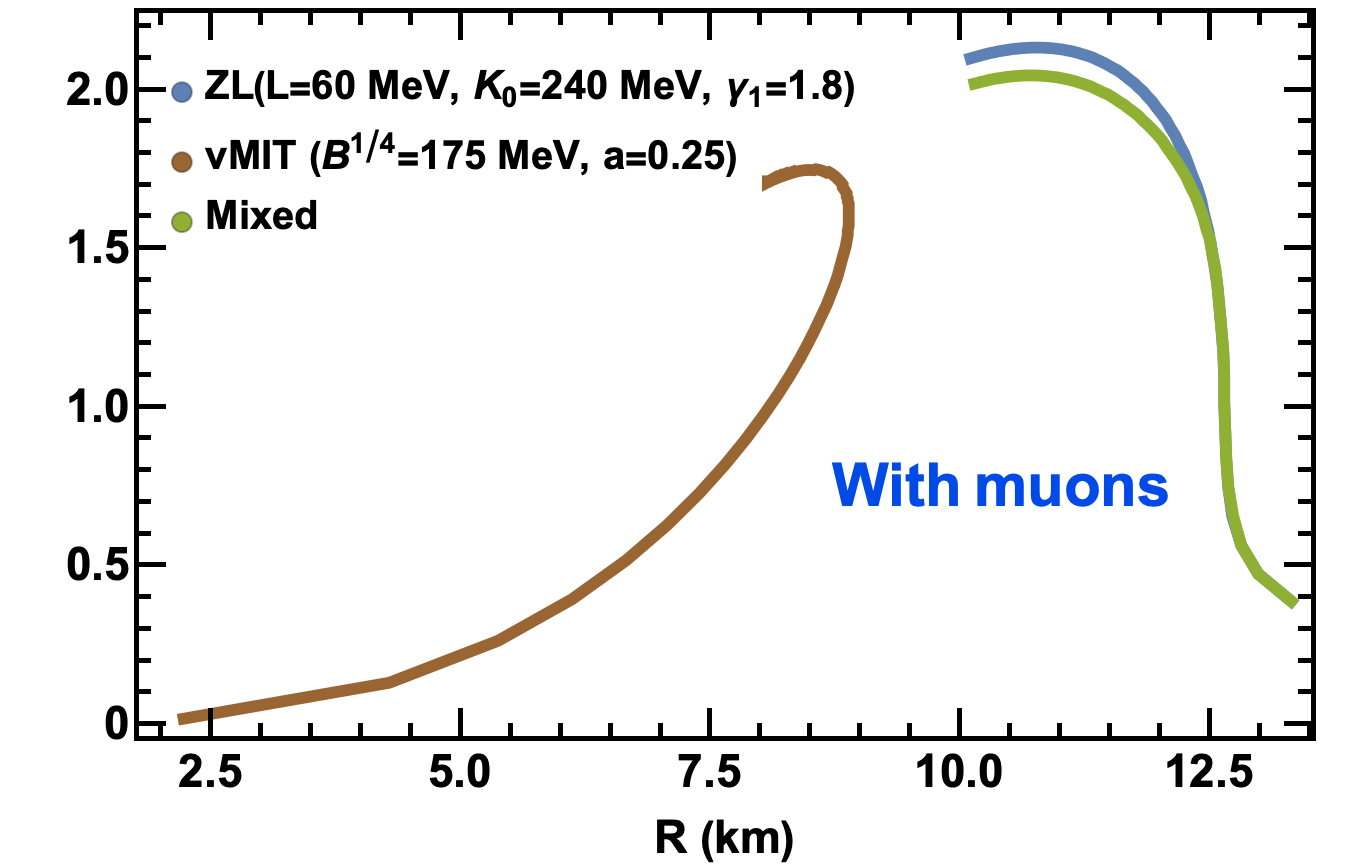}\\[-2ex]
}
\caption{\label{fig:8} The mass-radius curves for a hybrid star (Gibbs construction) with EOS parameters chosen such that the mixed phase supports 
$M_{\rm max}=2.05~M_\odot$.  In the left panel, muons are not included, whereas the right panel is with muons included.}
\label{MR-Mixed-Stiff}
\end{figure*}

Figure~\ref{MR-Mixed-Stiff} shows $M$-$R$ curves for a hybrid star whose $M_{\rm max} = 2.05~M_\odot$.  
This value 
is obtained 
by increasing the compressibility parameter of the ZL EOS from $K_0$ = 220 to $K_0$ = 240 MeV, and increasing $\gamma_1$ from 1.6 to 1.8 while maintaining causality. Including muons pushes the onset of the mixed phase to slightly higher densities, which causes the maximum mass of a hybrid star with muons to be higher than for a hybrid star without muons. This is in contrast to the effect of muons in an ordinary NS, where the softening results in a lower maximum mass. The leftmost curves in these figures refer to a self-bound quark star, and are shown here to provide contrast. 

\subsection{Boundary conditions for the $g$-mode oscillation}

Having established the equilibrium structure and computed the sound speeds, we have all the variables necessary to solve Eqs.~(\ref{oscr}) at hand, except for the boundary conditions that determine the (real) eigenfrequencies. The boundary conditions for Newtonian structure equations are obtained as a straightforward limiting case of Eqs. (\ref{oscr}), and are discussed at length in~\cite{Lai94}. To summarize those results, in the Newtonian non-relativistic case, regularity of $\xi_r,~\delta p/\rho$ can be checked by Taylor expansion around $r$ = 0. The resulting condition is:
\beq
r^2\xi_r=\frac{l}{\omega^2}(Y_0+\phi_0)r^{l+1},\quad \frac{\delta p}{\rho}=Y_0r^l \,,
\eeq
where $Y_0,~\phi_0$ are constants. For our purposes, $\phi_0$ = 0 since we ignore perturbations in the gravitational potential, as in~\cite{Lai94}. $Y_0$ is an arbitrary normalization constant allowed by the linearity of these equations. Effectively, this means that the overall scale of the eigenfunctions is arbitrary. It must be determined by external factors, such as the strength of the force (tidal effects in a merger, for example). The normalization has no impact on the numerical value of the eigenfrequency. It is therefore conventional to choose $Y_0$ = 1. We will make, for simplicity, and without loss of generality, a slightly different choice:
\begin{equation}
\frac{l}{\omega^2}Y_0=1
\end{equation}
so that (for $l$=2), $\xi_r\rightarrow r$ as $r\rightarrow 0$. In practice, we start the integration slight off-center, so $\xi_r$ will be small but non-zero. The other condition at the center becomes
\begin{equation}
    \frac{\delta p}{\rho}=\frac{\omega^2}{l}r^l{\rm e}^{-\nu_0} \,,
\end{equation}
again, with $l$ = 2 for our case. For the relativistic form of the oscillation equations, the above conditions still apply with the change $\frac{\delta p}{\rho}\rightarrow \frac{\delta p}{\epsilon+p}$. The boundary condition at the surface is the vanishing of the Lagrangian pressure perturbation $\Delta p=c_s^2 \Delta \epsilon=0$. This projects out the radial component of $\vec{\xi}$. In the non-relativistic case, $\nabla p=-\rho g$ while in the relativistic case, $\nabla p=-g h$ with $h=(\epsilon+p)$ the enthalpy. 

With some algebra, one can arrive at a simpler form of Eqs. (\ref{oscr}):
\begin{eqnarray}
\label{eq:uv}
\frac{dU}{dr}&=&\frac{g}{c_s^2}U+{\rm e}^{\lambda/2}\left[\frac{l(l+1){\rm e}^{\nu}}{\omega^2}-\frac{r^2}{c_s^2}\right]V \nonumber \\ 
\frac{dV}{dr}&=&{\rm e}^{\lambda/2-\nu}\frac{\omega^2-N^2}{r^2}U+g\Delta(c^{-2})V \,,
\end{eqnarray}
where $U$ = $r^2{\rm e}^{\lambda/2}\xi_r$, $V$ = $\delta p/(\epsilon+p)$ and  \\ $\Delta(c^{-2})=c_e^{-2}-c_s^{-2}$. 
 
 We employ a 4th-order Runge-Kutta scheme to find a global solution of the linear perturbation equations, Eqs. (\ref{eq:uv}), subject to the boundary conditions for the relativistic case outlined above. Since the solution set comprises overtones, we selected the lowest order $g$-mode (highest frequency) by checking that the radial eigenfunction $\xi_r$ has only one node inside the star. The corresponding eignefrequency is plotted in the figures that follow. We examine the trends of the $g$-mode vs. mass for various parameter choices, for the pure nuclear, self-bound and hybrid stars.

\begin{figure*}[htbp]
\parbox{0.48\hsize}{
\includegraphics[width=9cm,height=5.5cm]{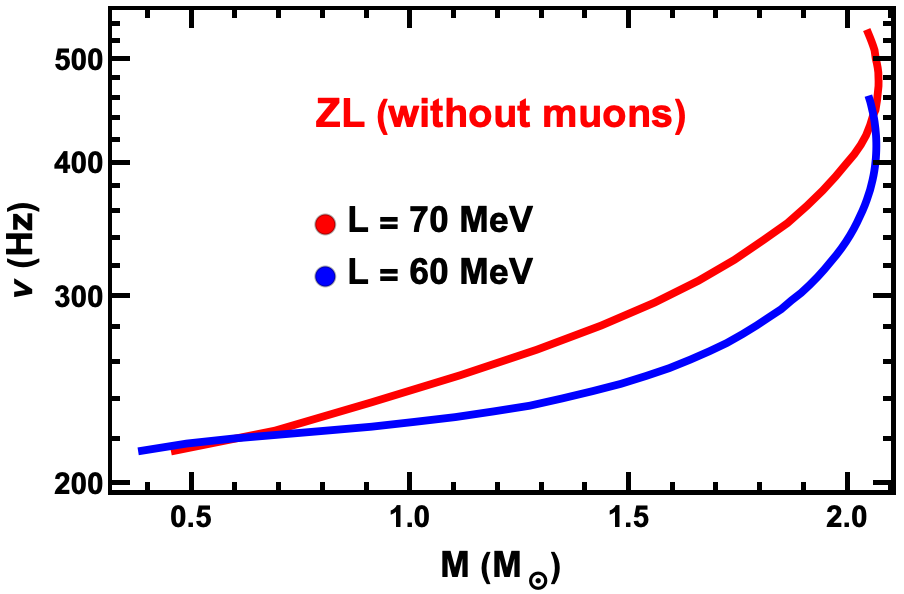}\\[-2ex]
}\parbox{0.54\hsize}{
\includegraphics[width=8.2cm,height=5.5cm]{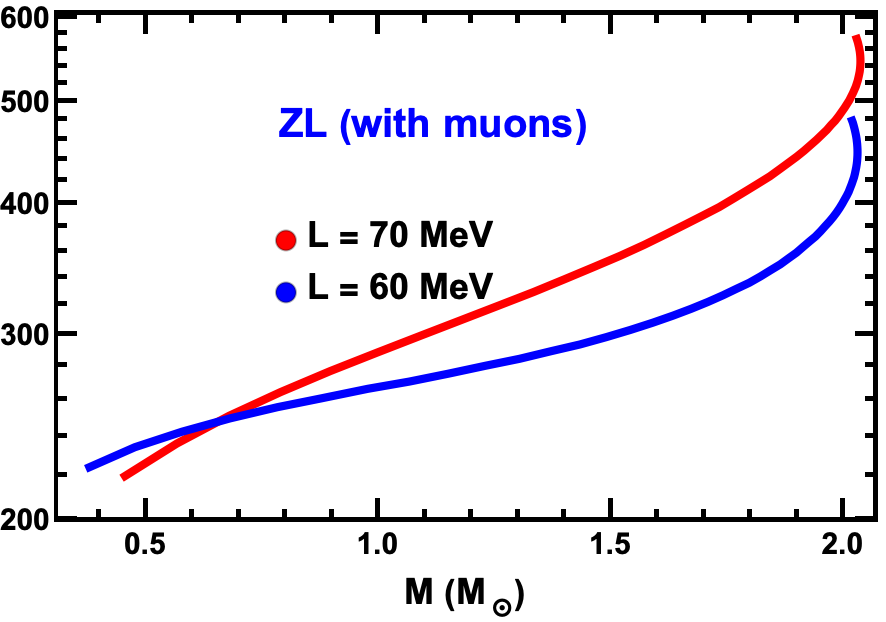}\\[-2ex]
}
\caption{Contrasts of the $g$-mode frequencies vs mass of  normal NSs for the ZL EOS  without and with muons. 
The two curves with different $L$'s in each panel are for EOSs with $K_0=220$ MeV, $S_v=31$ MeV and $\gamma_1$=1.6. 
}
\label{fig:g-mode-freq-ZLs}
\end{figure*}

Figure \ref{fig:g-mode-freq-ZLs} contrasts the influence of varying the density dependence of the symmetry energy, by changing the slope of the symmetry energy parameter $L$ at $n_s$, of the underlying ZL EOS for normal neutron stars with  fixed $K_0=220$ MeV and $S_v=31$ MeV.   
For $L$ = 60 MeV as well as $L$=70 MeV, the softening effect of muons leads to a noticeable increase in the $g$-mode frequency at a given mass. Comparing $L$ = 60 MeV with $L$ = 70 MeV for a fixed composition however, the $g$-mode frequency for $M\simgreater~$0.5-0.6 \msun is higher for the stiffer EOS.

\begin{figure*}[hbtp]
\parbox{0.48\hsize}{
\includegraphics[width=8.6cm,height=5.5cm]{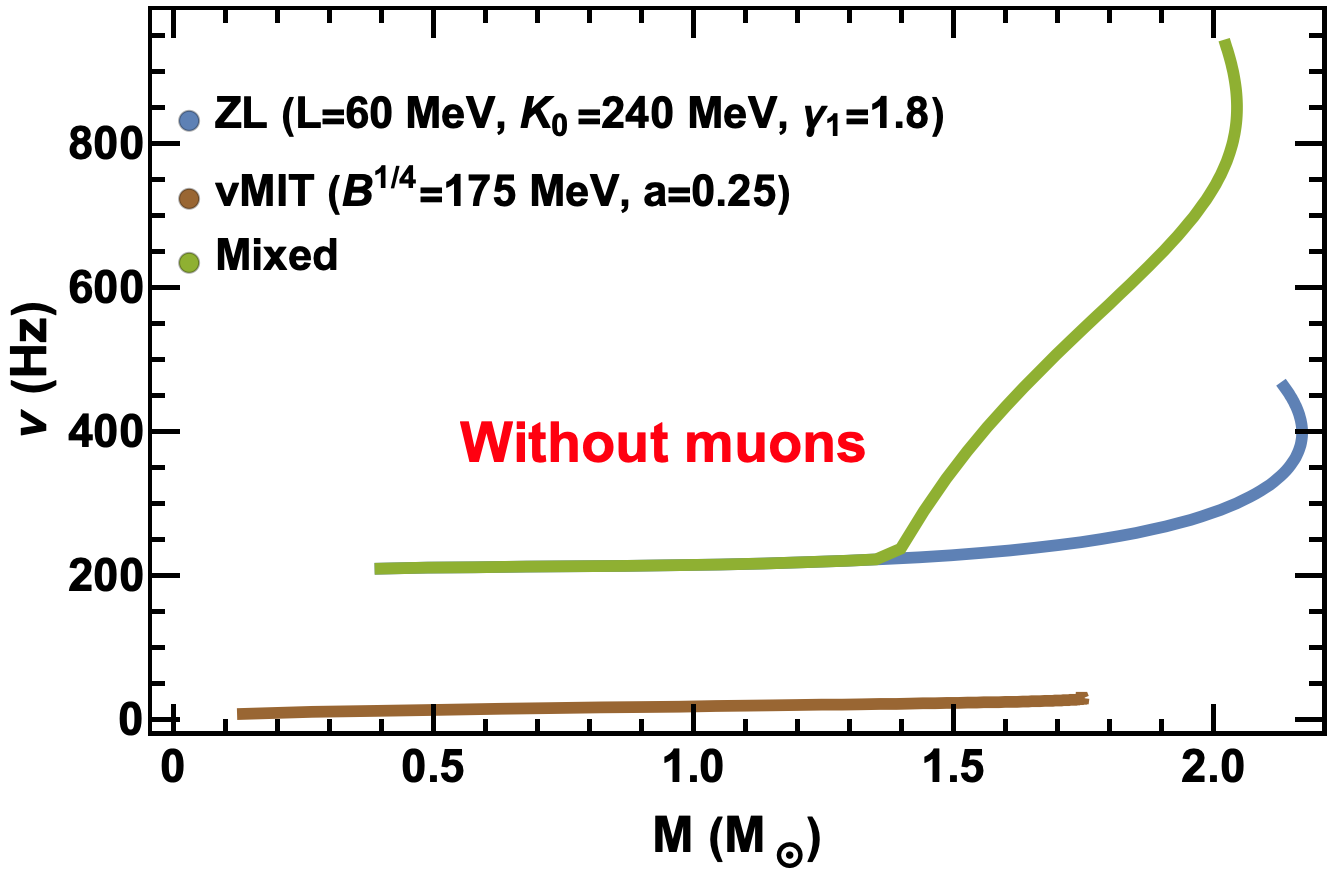}\\[-2ex]
}\parbox{0.48\hsize}{
\includegraphics[width=8.5cm,height=5.5cm]{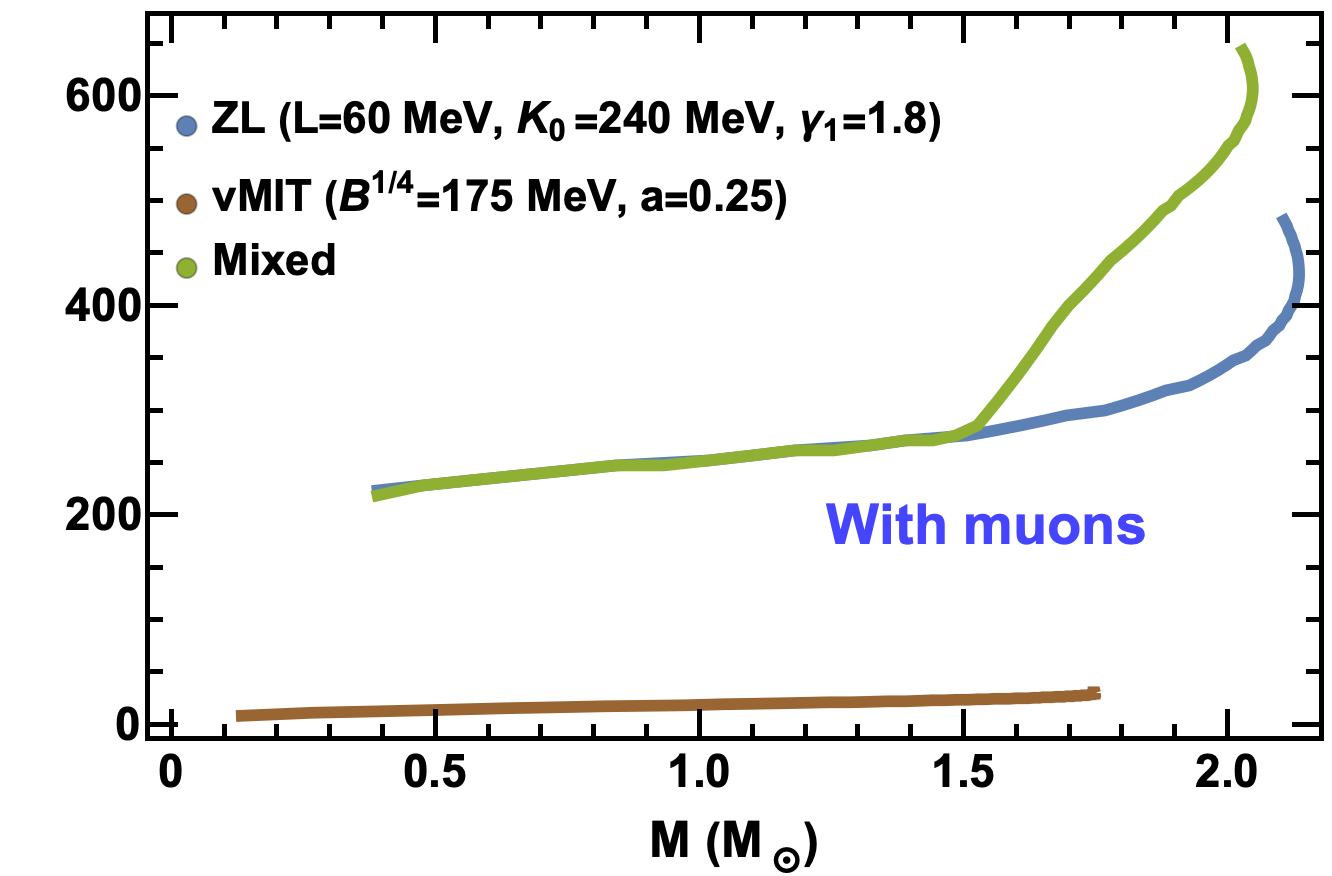}\\[-2ex]
}
\caption{Contrasts of g-mode frequencies vs stellar mass in a hybrid star. Parameters of the EOSs are as in the insets. In the left panel, muons are not included, whereas the right panel is with muons included.}
\label{g-mode-freq-Mixed}
\end{figure*}

In Fig.~\ref{g-mode-freq-Mixed}, results contrasting the $g$-mode frequencies in normal, hybrid, and self-bound stars are presented. The contents of this figure constitute the principal result of this work, viz., the abrupt rise in the scale of the $g$-mode frequency at the onset of the mixed phase in the hybrid star. For the EOS parameters displayed in the figure, the jump occurs around 1.4 $M_\odot$, so that a hybrid star in a merger would have a distinctly higher g-mode frequency than a normal NS. In the top panel, the ZL EOS does not include muons, whereas in the bottom panel the ZL EOS includes muons. The $g$-mode frequency in the mixed phase is again higher than in a pure phase, but since the mixed phase appears at a higher density due to muons, the rise in the $g$-mode is less dramatic compared to a hybrid star without muons. Results for the self-bound star are shown here for comparison, and to emphasize that its $g$-mode frequency is comparatively small (10-50 Hz). Unlike the $f$-mode frequency for the hybrid star, which gradually interpolates between those of the normal NS and self-bound star~\cite{Flores13,Flores:2017kte} and shows no dramatic effects of compositional changes, the $g$-mode frequency for the hybrid star is the highest of all and is sensitive to the onset of quarks - making it less subject to ambiguity. One does not need to know the mass of the star to ascertain if it can be a hybrid star if the $g$-mode frequency can be precisely determined. 

The unusually large \gm~frequency for the hybrid star with a Gibbs mixed phase may be understood in a qualitative sense using general thermodynamic considerations without reference to details of the EOS. In general, the equilibrium sound speed $c_{e,\mathrm{mix}}$ in a system with two conserved charges ($\mu_B$ and $\mu_Q$) can be expressed as
\begin{eqnarray}
    c_{e,\mathrm{mix}}^{2}&=&\frac{d p_{\mathrm{mix}}\left(\mu_{B}, \mu_{Q}\right)}{d \epsilon_{\mathrm{mix}}} \nonumber \\
    &=&\frac{\partial p_{\mathrm{mix}}}{\partial \mu_{B}}\left(\frac{d \mu_{B}}{d \epsilon_{\mathrm{mix}}}\right)+\frac{\partial p_{\mathrm{mix}}}{\partial \mu_{Q}}\left(\frac{d \mu_{Q}}{d \epsilon_{\mathrm{mix}}}\right)
    \label{Gibbs-sound}
\end{eqnarray}
where $\mu_Q$ is the charge chemical potential. Glendenning~\cite{Glen} showed that in such a situation, while $\mu_B$ is smooth at the onset of the mixed phase, $\mu_Q$ is not, as there is freedom to rearrange charges between the two phases to achieve global charge neutrality and minimize the free energy. In fact, the steady rise with density of $\mu_Q$ in the pure nuclear phase changes abruptly to a decline in the mixed phase, 
tempering the equilibrium sound speed as shown by our numerical results presented in Fig.~\ref{fig:c2diff-Mixed} and confirmed by other works~\cite{2019PhRvC..99f5802W} which use  different EOS from ours for the nucleon and quark sector. On the other hand, the adiabatic sound speed $c_{s,\mathrm{mix}}$ is evaluated at fixed composition and shows no such effect, hence the difference of the two sound speeds (usually small in a pure phase) abruptly increases in the mixed phase. This is reflected as a positive jump in the \bv~frequency and therefore of the \gm~ in the mixed phase.
 
\section{$g$-mode Energy and Tidal Forcing}
\label{sec6}
Unlike the Sun, where convection from within can drive oscillations, any oscillations of an evolved NS likely require an external agent to excite and sustain the perturbation beyond its normal damping time. A violent event such as a NS merger is bound to produce oscillations in the pre-merger phase due to tidal forcing or in the postmerger (ringdown) phase as the hypermassive remnant relaxes to its stable rotating configuration. Here, we estimate the impact of the $g$-mode on tidal phasing leading up to the merger, as the $g$-mode spectrum in the postmerger remnant can be modified by thermal and convective effects which are beyond the scope of the current work. We follow 
\cite{RG2} and  
assume spherically symmetric non-rotating stars, the Newtonian approximation to orbital dynamics and quadrupolar gravitational wave emission. These simplifying approximations allow for a first estimate of the excitation energy and amplitude of the $g$-mode, as well as the phase difference due to dynamic tides associated to the $g$-mode (not to be confused with the quasi-static tides due to global deformation). Our estimates can be systematically improved by going to the post-Newtonian approximation or numerical relativity.

\indent The estimates are derived by modeling the NS as a forced simple harmonic oscillator with a mass $M_{\ast}$, radius $R_{\ast}$ and a natural frequency $\omega$=$\omega_g$, the angular frequency of the $g$-mode. The forcing comes from the quadrupolar moment of the companion star's gravitational force (mass $M$), which couples to the $g$-mode. By following the analysis of~\cite{RG2}, we arrive at an expression for the accumulated phase error $\Delta \Phi(t)$ caused by the $g$-mode:
\beq
\Delta \Phi(t) \approx \frac{3 \pi \Gamma}{m}\left[\frac{\Omega_{e}(t)}{\Omega}-1\right]\left(-\frac{\Delta E}{E}\right) \,,
\label{phieqn}
\eeq
where $\triangle E$ is the energy pumped into the $g$-mode, $E$ the  total (kinetic plus potential) orbital energy of the system, 
$\Omega_{e}(t)$ the time-dependent orbital frequency of the binary, and $\Omega$=$\omega_g/m$. 
The quantity $m$ in Eq. (\ref{phieqn}) is the azimuthal mode index ($m$=2 in this case). Finally, $\Gamma$ is a quantity that appears as a result of applying the stationary phase approximation to the evaluation of the time to resonance~\cite{RG2}, and is quantified below. A $\Delta \Phi(t)$ of ${\cal O}(1)$ signifies a large deviation from the point particle approximation to the gravitational waveform from the merger. Explicitly, the quantity $\triangle E$ 
(for  
angular quantum number $l$) is given by
\begin{eqnarray}
\Delta E &=&\left(\frac{5\pi}{384m}\right)\frac{M/M_{\ast}}{[1+M/M_{\ast}]^{(2l+1)/ 3}}\left(\frac{c^{2} R_{\ast}}{G M_{\ast}}\right)^{5 / 2} 
\nonumber \\
 &\times& \left(\frac{\Omega}{\Omega_d}\right)^{(4 l-7) / 3}\left(\frac{G M_{\ast}^{2}}{R_{\ast}}\right) S_{lm}^{2} \,.
\label{delE}
\end{eqnarray}
where $\Omega_d$=$(G M_{\ast}/R_{\ast}^3)^{1/2}$ is a 
natural frequency unit and $S_{lm}$ is proportional to the overlap integral between the mode eigenstate $|lm\rangle$ and the vector spherical harmonic
$\left|P_{l m}\right\rangle=\nabla\left[r^{l} Y_{l m}(\theta, \phi)\right]$. The total instantaneous orbital energy is $E=-GMM_{\ast}/2a$ with $a=a(t)$ the instantaneous orbital separation.
The evolution of the orbital frequency for a circularized orbit using the formula for quadrupolar gravitational wave emission gives
\beq
\Omega_{e}(\tau)=\frac{1}{8}\left(\frac{a M_{c}}{c^{3}}\right)^{-5 / 8} \frac{1}{\tau^{3 / 8}} \,,
\label{orbit-evo}
\eeq
where $M_c$ is the chirp mass of the binary system and $\tau$ is the time to coalescence. All quantities on the right hand side in Eq.~(\ref{phieqn}) can be calculated, once the parameters of the binary ($M, M_{\ast},R_{\ast}$) and the resonant \gm~frequency are fixed. We choose $M$=$M_{\ast}$=1.5 $M_{\odot}$ for neutron/hybrid stars and pure quark stars. The strongest tidal coupling is likely to the $l$=$m$=2 $g$-mode whose characteristic frequency we choose as 
\begin{eqnarray}
\omega_{g}(N S)  &\cong&  2\pi(200)\, {\rm Hz}\,, ~~ \nonumber \\
\omega_{g}(H S)   &\cong& 2\pi(300)\, {\rm Hz}\,, ~~ \rm{and} ~~ \nonumber \\
\omega_{g}(Q S)   &\cong& 2\pi(40)\, {\rm Hz} \,, ~~ 
\end{eqnarray}
based on 
the $g$-mode eigenfrequencies 
in the previous section. Even without computing $S_{lm}$, one can estimate from Eq. (\ref{orbit-evo}) the time at which the $g$-mode becomes resonant as 
\begin{eqnarray}
\tau_{0}(N S)  &\cong&  272\, {\rm ms}\,, ~~ \nonumber \\
\tau_{0}(H S)  &\cong&  103\, {\rm ms} \,, ~~ \rm{and} ~~ \nonumber \\
\tau_{0}(Q S)  &\cong& 22\, {\rm s} \,,
\end{eqnarray}
where the the zero of time is the moment of coalescence.
%
%
Assuming circularized orbits, standard equations of
binary orbit evolution $a(t)$~\cite{MM} give
\begin{eqnarray}
a_{0}(NS) &\cong& 111~ \mathrm{km} \,, ~~ \nonumber \\
a_{0}(HS) &\cong& 85~ \mathrm{km} \,, ~~ {\rm and} ~~ \nonumber \\
a_{0}(QS) &\cong& 326~ \mathrm{km} \,.
\end{eqnarray}
%
%
%
We note that the $g$-mode for the hybrid star, which has a larger resonant frequency than neutron or quark stars, is excited later in the merger and is likely to be stronger in amplitude due to the close separation of the binary since the forcing term is $\propto 1/a^3$ for $l$=2. 
Finally, from our calculations for the $g$-mode eigenfunction and the associated density perturbation $\delta\epsilon(r)$, we estimate
\begin{eqnarray}
S_{lm}^{NS} &\cong& \quad 4.5 \times 10^{-3} \,, ~~ \nonumber \\
S_{{lm}}^{HS} &\cong& \quad 6.2 \times 10^{-3} \,~~ {\rm and} ~~ \nonumber \\
S_{lm}^{QS} &\cong& \quad 9.9 \times 10^{-6} \,
\end{eqnarray}
using Eq. (40) for $S_{lm}$ from~\cite{Yu17}.
%
From these estimates, Eq. (\ref{delE}) can be utilized to yield the estimated fractional orbital energy pumped into the $g$-mode:
\begin{eqnarray}
\left|\frac{\Delta E}{E}\right|^{N S} &\cong& 2.3 \times 10^{-3} \,,  ~~ \nonumber \\
\left|\frac{\Delta E}{E}\right|^{H S } &\cong& 5.9 \times 10^{-3} \,, ~~ \nonumber ~~ {\rm and} ~~\\
\left|\frac{\Delta E}{E}\right|^{Q S} &\cong& \quad 2 \times 10^{-9} \,, 
\end{eqnarray}
%
%
and finally from Eq. (\ref{phieqn}), we obtain the phase error due to the resonant excitation of the $g$-mode to be
\begin{eqnarray}
\triangle \phi^{NS} &\cong& 0.8 \,, ~~ {\rm and} ~~ \nonumber \\
\triangle \phi^{HS} &\cong& 0.45  \,, ~~ \nonumber \\
\triangle \phi^{QS} &\cong& 6 \times 10^{-4} \,.
\end{eqnarray}
%
%
%
Note that $\Delta \phi^{NS}$ and $\triangle \phi^{HS}$ are comparable. Despite $\left(\frac{\Delta E}{E}\right)$ being larger for a hybrid star as expected, its higher $g$-mode frequency means it is excited later in the merger, when there is less time left for accumulating a phase error. These results are very sensitive to the value of $S_{lm}$ ($\Delta\Phi \propto S_{lm}^{2}$), which itself can vary by a factor of 2 or more depending on the EOS. 

\subsection*{Comparison with other works}


\begin{table}[htbp!]
\caption{Comparison of characteristic $g$-mode frequencies (denoted by $\omega_g$ in the table) reported in a selection of the literature. As other works usually fix the stellar mass $M$, we include this information. The symbol $\Lambda$ is used here as a shorthand to denote hyperonic degrees of freedom and SF denotes superfluidity in the nucleonic sector. 
Values of $f_g$ that vary with the NS mass can be inferred from  Figs. \ref{fig:g-mode-freq-ZLs} and \ref{g-mode-freq-Mixed} of this work. 
The entries are representative, not exhaustive.} 

\begin{center} 
\begin{tabular}{cccc}
\hline
\hline
Authors~[Ref.] & Core &  $M$ & $f_g=\omega_g/(2\pi)$  \\
& Composition & [$M_\odot$] & [kHz]  \\ \hline
Reisenegger \& Goldreich~\cite{RG} & $npe$ & 1.405 & 0.215  \\
Lai~\cite{Lai94} & $npe$ & 1.4 & 0.073  \\
Kantor \& Gusakov~\cite{Kantor14} & $npe$ & 1.4 & 0.13 \\
Kantor \& Gusakov~\cite{Kantor14} & $npe\mu$ & 1.4 & 0.19 \\
Kantor \& Gusakov~\cite{Kantor14} & $npe\mu$(SF) & 1.4 & 0.46 \\
Dommes \& Gusakov~\cite{Dommes15} &  $npe\mu\Lambda$(SF) & 1.634 & 0.742 \\
Yu \& Weinberg~\cite{Yu-Wein} & $npe\mu$ & 1.4 & 0.13 \\
Yu \& Weinberg~\cite{Yu-Wein} & $npe\mu$(SF) & 2.0 & 0.45 \\
Rau \& Wasserman~\cite{Rau18} & $npe\mu$(SF) & 2.0 & 0.45  \\\\
Jaikumar~ et al. [this work] & $npe$ & 1.4 & 0.24  \\
Jaikumar~ et al. [this work] & $npe\mu$ & 1.4 & 0.27  \\
Jaikumar~ et al. [this work] & $npe\mu q$ & 2.0 & 0.58  \\
\hline \hline
\end{tabular}
\end{center}
\label{Comparison}
\end{table}

 Table~\ref{Comparison} compares our results for zero-temperature core $g$-modes in the Gibbs mixed phase of hadrons and quarks with other works, some of which also find an enhancement of the frequency due to other compositional mixes or collective fluid effects like superfluidity, although values in the table do not include the effect of entrainment on the $g$-mode, which has also been studied. Details about the different EOSs used, the effect of non-zero temperature and entrainment can be found by perusing the respective reference. We confirm the value of the $g$-mode frequency for $npe$ and $npe\mu$ non-superfluid matter described by the Akmal-Pandharipande-Ravenhall (APR)-EOS as reported in~\cite{Kantor14}, which also serves as a test of our numerics. In comparison to ~\cite{Kantor14} with the APR-EOS or ~\cite{Yu-Wein} with the SLy4 equation of state, the ZL-EOS yields a larger value of the $g$-mode frequency as it is less stiff than either of those two. While the EOS and the treatment of gravitational perturbations differ between the cited works, the results for $npe\mu$ matter with superfluidity appear to be in general agreement with each other. Note the considerably larger value of the $g$-mode frequency for hyperonic stars with superfluidity compared to hybrid stars or superfluid neutron stars. All of these, in turn, are larger than non-superfluid neutron/hyperonic stars although the latter employ Newtonian gravity. A study of $g$-mode frequencies and damping times in superfluid hybrid stars is a future objective that would make this comparison more complete.

\section{Summary and Conclusions}
\label{sec7}

The main objective of this work was to ascertain the characteristics of $g$-mode oscillations of NSs containing 
QM in their cores. Toward this end, the nucleon-to-quark phase transition was treated using Gibbs construction which renders an otherwise sharp first order transition smooth. 
The cores of such hybrid stars accommodate admixtures of nucleonic and quark matter, the pure quark phase being never achieved. This feature, while allowing contrasts between the structural properties (e.g., $M$ vs $R$) of normal and hybrid stars to be made also permits comparisons of observables that depend on their interior compositions, such as short- and long-term cooling, oscillation modes, etc. Determining the composition of the star is essential to break the degeneracy that exists in the masses and radii 
 of normal and hybrid stars as one may be masquerading as the other.    

The nucleonic part of the EOS used in this work tallies with nuclear systematics near and below $n_s$ in addition to being  consistent with results from modern chiral EFT calculations up to $2n_s$ for which  uncertainty quantifications have been made. The EOS employed in the quark sector is sufficiently stiff, hence non-perturbative, to support $\sim 2M_\odot$ NSs  required by recent observations. Furthermore, the overall EOS gives radii of $\sim 1.4M_\odot$  stars that lie within the bounds of recent determinations. The EOS is also consistent with the tidal deformation inferred from gravitational wave detection in the event GW170817. Appendix A summarizes the structural properties for the EOSs used and provides mathematical details for the derivation of the sound speeds.

Unlike for $M$-$R$ curves for which only the pressure vs density relation (EOS) is sufficient,  the analysis of 
\gm~oscillations requires simultaneous information about the equilibrium and adiabatic squared sound speeds, $c_e^2=dp/d\epsilon$ and 
$c_s^2=\partial p/\partial\epsilon|_x$, where $x$ is the local  proton fraction.  The distinction between these two sound speeds plays a central role in determining the \bv~frequencies $\omega^{2} \propto c_e^{-2} - c_s^{-2}$ of non-radial \gm~oscillations. 
Thus, a future detection of \gms~would take gravitational wave astronomy beyond the current capability of $M$-$R$ measurements to determine the composition of the star.

We find that the $g$-mode is sensitive to the presence of QM in NSs, where 
quarks are 
part of a mixed phase with nuclear matter in the core. The equilibrium sound speed drops sharply at the boundary of the mixed phase (Fig. 5), raising the local \bv\, frequency and the fundamental \gm\, frequency of the star (Fig. 6). Contrasts of \gm\ frequencies between normal and hybrid stars containing quark matter (Fig. 9) form the principal results of our work.

Our analysis leads to the conclusion that in binary mergers where one or both components may be a hybrid star, the fraction of tidal energy pumped into the resonant \gm ~in hybrid stars can exceed that of a NS by a factor of 2-3, although resonance occurs later in the inspiral. On the other hand, a self-bound star has a much weaker tidal overlap with the g-mode. The cumulative tidal phase error in hybrid stars  $\Delta\phi\cong$ 0.5 is comparable to that from tides in ordinary NSs. While this happenstance may present a challenge in distinguishing between the two cases, should the g-mode be excited to sufficient amplitude  in a postmerger remnant, its frequency spectrum would be a possible indication for the existence of non-nucleonic matter, including quarks. The detection of such \gm~frequencies in binary mergers observed by current gravitational wave detectors seems challenging, but possible with next generation detectors. 

The novel features of this work include (i) use of nucleonic EOSs that are consistent with constraints from modern chiral EFT calculations coupled with sufficiently stiff quark EOSs to calculate structural properties of hybrid stars that lie within the bounds of astrophysical measurements, (ii) a first calculation of the two sound speeds and the principal $g$-mode frequency of hybrid stars employing Gibbs phase criteria, and (iii) a concomitant analysis of tidal phase effects in a binary merger due to $g$-modes in hybrid stars.
In  future work, we aim to report on $g$-mode frequencies in alternative treatments of quark matter in NSs such as a first order nucleon-to-quark phase transition and crossover transitions  as in quarkyonic matter. \\

\emph{Acknowledgments.}--- We are grateful to the anonymous referee for meticulous review of the equations. We 
acknowledge discussions with Thomas Kl\"{a}hn on the non-perturbative EOS for quark matter. Thanks are due to Sophia Han for remarks on observational constraints on the EOS.  P.J. is supported by the U.S. NSF Grant No. PHY-1608959 and PHY-1913693. The research of A.S. and M.P. was supported by the U.S. Department of Energy, Grant No. DE-FG02-93ER-40756. 
C.C. acknowledges support from the European Union’s Horizon 2020 research and innovation programme under the Marie Sk\l{}odowska-Curie grant agreement No. 754496 (H2020-MSCA-COFUND-2016 FELLINI).


\appendix

\section{Determination of EOS constants in SNM and PNM for the ZL EOS}
\subsection{\label{app:subsec3}SNM}

The constants $a_0, b_0$ and $\gamma$ in Eq.~(\ref{asym}) for SNM are determined by utilizing the empirical properties of SNM at $u=1$. 
Specifically, the values used are $E_{1/2} = -B = -16$ MeV at  $n_s = 0.16~{\rm fm^{-3}}$, $p_{1/2}/n_s=0$, and $K_{1/2} =220$ MeV. Manipulating the relations
\beq
- B &=& T_{1/2} + a_0 + b_0  \\
0 &=& T_{1/2}^\prime + a_0 + \gamma b_0 \\
\frac{K_{1/2}}{9} &=&   T_{1/2}^{\prime\prime} + 2 T_{1/2}^{\prime} + 2 a_0 + \gamma (\gamma+1) b_0 \,,  
\eeq
the constants are given by 
\begin{eqnarray}
\label{aconst}
\gamma &=& \frac {K_{1/2}/9 - T_{1/2}^{\prime\prime}} {T_{1/2} - T_{1/2}^{\prime} +B } \,, \quad \nonumber \\
b_0 &=& \frac {K_{1/2}/9 - T_{1/2}^{\prime\prime}} {\gamma (\gamma-1)} \quad {\rm and} \nonumber \\
a_0 &=& - B - T_{1/2} - b_0 \,,
\end{eqnarray}
where $ T_{1/2}^\prime = u \frac{dT_{1/2}}{du}$ and $T_{1/2}^{\prime\prime} = u^2 \frac {d^2T_{1/2}}{du^2}$.   
Explicit expressions for these derivatives are
\beq
\label{akinenergy}
 T_{1/2}^\prime &=& \left. \frac {p_{1/2}^{\rm kin}} {n}\right|_{n_s} \nonumber \\
 &=& \frac {1}{n_s} \cdot \frac {2}{12\pi^2} 
 \bigg[ k_F E_F \left(  k_F^2 - \frac 32 M_B^2 \right) \nonumber \\
 &+& \frac 32 M_B^4 \ln \left( \frac {k_F+E_F}{M_B} \right)  \bigg]_{k_{Fs}} \,, 
 \nonumber \\
 T_{1/2}^{\prime\prime} &=& \frac{K_{1/2}^{\rm kin}}{9} - 2 T_{/12}^\prime = \frac {k_{Fs}^2}{3E_{Fs}} - 2 T_{/12}^\prime ,
\eeq
 where $k_{Fs}=(3\pi^2n_s/2)^{1/3}$. To obtain the first term in the rightmost equality above, it is advantageous to use the thermodynamical identity $p=n\mu-\epsilon$ for the kinetic parts, whence $\frac {dp}{dn} = n \frac {d\mu}{dn} = \frac {d\mu}{dk_F} \frac {dk_F}{dn}$. The result quoted above ensues from the relations   $\frac {dk_F}{dn} = \frac {k_F}{3n}$ and $\mu=E_F=\sqrt{k_F^2 + M_B^2}$ both evaluated at $n_s$ and $k_{Fs}$. 

Numerical values of the derivatives and constants so derived are 
\begin{eqnarray}
T_{1/2} &\simeq& 21.79~ {\rm MeV}, ~~~  T_{1/2}^\prime \simeq 14.34~{\rm  MeV,} \nonumber \\
T_{1/2}^{\prime\prime} &=&  -5.030~{\rm MeV,} ~~~\gamma \simeq 1.256, \nonumber \\ 
~~~a_0 &\simeq& -129.3~{\rm MeV}, \quad {\rm and} ~~~b_0 \simeq 91.49~{\rm MeV}.
\end{eqnarray}
 as in ZL. For other permissible values of $K_{1/2}$ in the range $220\pm 30$ MeV, Eqs.~(\ref{aconst}, \ref{akinenergy}) can be used to determine the corresponding constants. \\

\subsection{\label{app:subsec4}PNM}

In the PNM sector in which $x=0$, the constants in Eq.~(\ref{EPNM}) to determined are $a_1,b_1$ and $\gamma_1$. As in SNM, $E_0$ and $T_0$ are relative to the baryon mass $M_B$.  Denoting the energy per baryon of PNM by $E_0$, its various terms and the associated pressure are
\begin{eqnarray}
E_0 &=& T_0 + V_0 =  T_0 + a_1 u + b_1 u^\gamma_1 \,, \nonumber \\
p_0 &=& n_s \left(u^2 \frac {dE_0}{du} \right) \nonumber \\
&=& n_s \left( u^2 T_0^\prime + a_1 u^2 + \gamma_1 b_1 u^{\gamma_1+1} \right)\,. 
\end{eqnarray}
Evaluating the above equations at $u=1$ leads to
\beq
\label{E0}
E_0 = S_v - B = T_0 + a_1 + b_1 \,,  \\
p_0 = n_s (T_0^\prime + a_1 + \gamma_1 b_1) \,,
\label{pnmeqs}
\eeq
where $S_v = (E_0 - E_{1/2})$ at $u=1$. The last equation above is generally written as 
\beq
\frac {p_0}{n_s} &=& \frac L3 \quad {\rm with} \quad L = 3 \left[ n \frac {dS_v}{dn} \right]_{n_s} = 3~ [uS_v^\prime]_{u=1}  \quad {\rm so~that} \nonumber \\
\frac L3 &=& T_0^\prime + a_1 + \gamma_1 b_1 \,,
\label{Lo3}
\eeq
where $S_v^\prime = \frac {dS_v}{du}$. Manipulating Eqs. (\ref{E0}) and (\ref{Lo3}) leads to the relations
\begin{eqnarray}
b_1 &=& \frac {\frac L3 + B -S_v + T_0 - T_0^\prime} {\gamma_1 - 1} \quad {\rm and} \quad \nonumber \\
a_1 &=& S_v  - B - T_0 - b_1 \,.      
\end{eqnarray}
Taking guidance from the empirical properties of isospin asymmetric nuclear matter, we choose $S_v = 31$ MeV, $L$ in the range ($30$-$70$) MeV, and $\gamma_1=5/3$.  The resulting values of the constants are
\begin{eqnarray}
a_1 &\simeq& - \left( \frac L2 +14.72  \right)~{\rm MeV} \quad {\rm and} \quad \nonumber \\
b_1 &\simeq& \left( \frac L2 - 4.62 \right)~{\rm MeV}. 
\end{eqnarray}

\subsection{Sensitivity of the EOS constants}

The EOS constants above depend on the input values of 
$B,~n_s,~K_0$ and $S_v,~L,~\gamma_1$ in SNM and PNM, respectively.  Although we have used representative values for 
these quantities at $u=1$, nuclear data permits variations in them. 
Furthermore, one or more sets of these constants may be correlated, as for example, $S_V$ and $L$.  Additional constraints are to support $\simeq 2M_\odot$ NSs and to maintain causality, at least within the star.  These points must be borne in mind when varying the input values, particularly when correlated errors are present in theoretical evaluations of these quantities. \\


\subsection{NS properties with the ZL EOSs}

The various properties shown in Table~\ref{Struc2} below are for beta-equilibrated normal NSs  and correspond to variations in the characteristic properties of the ZL EOSs. \\


\begin{table}[htbp]
\caption{Structural properties of nucleonic NSs with $M=1.4\,M_\odot$ and $M_{\rm max}$ for 
the ZL EOSs. For each mass, the compactness parameter
$\beta=(GM/c^2R) \simeq (1.475/R)(M/M_\odot)$, $n_c$, $p_c$ and $y_c$ are the central values of 
the density,  pressure and proton fraction, respectively. The corresponding equilibrium squared speeds of sound 
are denoted by $c_e^2$. The $\Lambda$'s denote tidal deformabilities. } 

\begin{center} 
\begin{tabular}{crrrc}
\hline
\hline
Property              & ZL-A   & ZL-B & ZL-C   & Units \\ \hline
$K_0$  & 220 & 220 & 240 & {\rm MeV} \\
$S_v$ & 31 & 31 & 31 & {\rm MeV} \\
$L$  & 50 & 70 &  60 & {\rm MeV} \\ 
$\gamma_1$ & 1.6 & 1.6 & 1.8 \\
\hline 
$R_{1.4}$            & 11.77  & 12.69  & 12.61 & {\rm km} \\
$\beta_{1.4}$        & 0.175  & 0.163  & 0.164 & \\
$n_{c,1.4}/n_s$        & 3.35   & 2.78   & 2.75 & \\
$p_{c,1.4}$           & 83.65  & 60.59  & 61.50 & ${\rm MeV~fm^{-3}}$ \\
$(c_e^2)_{c,1.4}$     & 0.385  & 0.345  & 0.363 & $c^2$ \\
$\Lambda_{1.4}$ & 713.4 & 970.2 & 504.1 & \\
\hline
$R_{\rm max}$         & 10.01  & 10.68  & 10.8  & {\rm km} \\
$M_{\rm max}$        & 1.997   & 2.02   & 2.13 & $M_\odot$ \\ 
$\beta_{\rm max}$     & 0.294  & 0.279  & 0.291 & \\
$n_{\rm c,max}/n_s$    & 7.71   & 6.96   & 6.67 & \\
$p_{\rm c,max}$      & 798.89 & 602.04 & 646.7 & ${\rm MeV~fm^{-3}}$ \\
$(c_e^2)_{\rm c,max}$     & 0.874 & 0.777 & 0.866 & $c^2$ \\
$\Lambda_{\rm max}$ & 9.11 & 9.6 & 6.39 & \\
\hline \hline
\end{tabular}
\end{center}
\label{Struc2}
\end{table}

Structural properties of hybrid stars are discussed and shown in various figures in the text.

\subsection{Proof of equivalence of Eqs. (\ref{cs2du}) and (\ref{dc22})
}\vskip 0.2cm

Here we establish the equivalence of the direct approach to computing $c_s^2$ from Eq.~(\ref{cs2du})  with 
that from Eq.~(\ref{dc22}) for a general  
EOS with a parabolic dependence of the proton fraction $x$ in the case of $n,p,e$ matter. 
Both 
approaches 
yield identical results, which we have verified numerically as well. 

The equilibrium squared sound speed 
is
\begin{equation}
c_{e}^{2}=\left(\frac{d P}{d \epsilon}\right)_{\rm eq} =\frac{n_B\frac{d}{d n_{B}}\left(P_{\rm bar}+P_{e}\right)}{\left(\epsilon+P\right)} \,,
\end{equation}
 where the total pressure $P=P_{\rm bar}+P_e$ 
 is comprised of 
 the pressure from baryons and electrons. Writing 
 \begin{equation}
 \frac{d}{dn_B}=\frac{\partial}{\partial n_B}+\frac{dx}{dn_B}\frac{\partial}{\partial x} \,,
\end{equation}
 we get 
 \begin{equation}
c_e^2=\frac{\left(n_{B} \frac{\partial P_{b a r}}{\partial n_{B}}+n_{B} \frac{\partial P_{e}}{\partial n_{B}}\right)}{\epsilon+P} + \frac{n_{B}\left(\frac{\partial P_{\text {bar }}}{\partial x} \frac{d x}{d n_B}+\frac{\partial P_{\text {e}}}{\partial x} \frac{d x}{d n_B}\right)}{\epsilon+P}
\,.
\end{equation}
Comparing with Eq.~(\ref{cspd}), the first term on the right hand side is simply $c_s^2$. For the specific case of an 
EOS with parabolic dependence in $x$ 
(of which the APR-EOS in Kantor \& Gusakov~\cite{Kantor14} and the ZL-EOS~ in Zhao and Lattimer~\cite{zl2020} are examples), we have
\begin{equation}
    E(u,x)=E_0(u)+(1-2x)^2S_2(u);\quad u=n_B/n_0 \,.
    \label{EoS}
\end{equation}
%
where $n_0$ is the saturation density, $E_0$ the energy per baryons (neutrons and protons) and $S_2$ the symmetry energy. Computing the pressure and its derivatives with respect to $n_B$ and $x$ for the EOS in Eq.~(\ref{EoS}), we find 
\begin{equation}
 \frac{\partial P_{\text{bar}}}{\partial x}=-4 n_{B}(1-2x)uS_{2}^{\prime} \,,
\end{equation}
where the prime on $S_2$ is with respect to $u$. Since (assuming massless electrons) $\frac{\partial P_{\text {e}}}{\partial x}=\frac{1}{3} n_{B}\mu_{e}$, it follows that
 \begin{equation}
c_e^2=c_s^2 + \frac{1}{\mu_n}\left(\frac{\mu_{e}}{3}-4(1-2x)uS_{2}^{\prime}\right)n_B\frac{dx}{dn_B} \,.
\label{ce-again}
\end{equation}
From the $\beta$-equilibrium condition 
 \begin{equation}
\mu_e=4S_2(1-2x) \Rightarrow \mu_e/(4S_2)=(1-2x) \,,
\end{equation}
we get upon differentiation
\begin{equation}
\frac{1}{4 S_{2}}\left(\frac{d \mu_{e}}{d n_{B}}\right)-\left(\frac{\mu_{e}}{4 S_{2}^{2}}\right)\left(\frac{d S_{2}}{d n_{B}}\right)=-2\left(\frac{d x}{d n_{B}}\right) \,.
\label{x-eq}
\end{equation}
Using
\begin{equation}
\left(\frac{d\mu_e}{d n_{B}}\right)=\frac{\mu_{e}}{3 x}\left(\frac{d x}{d n_{B}}\right)+\frac{\mu_{e}}{3 n_{B}} 
\,,
\end{equation}
and solving for $dx/dn_B$ from Eq.(\ref{x-eq}), a minor rearrangement yields
\begin{equation}
n_B\left(\frac{d x}{d n_B}\right)=\frac{-\left(\frac{\mu_e}{3}-4(1-2 x) u S_{2}^{\prime}\right)}{\left(\frac{\mu_{e}}{3x}+8 S_{2}\right)}
\,.
\label{xsol}
\end{equation}
Putting together Eq.~(\ref{xsol}) with Eq.~(\ref{ce-again}), we get 
\begin{eqnarray}
c_e^2=c_s^2 - \frac{1}{\mu_n}\frac{\left(\frac{\mu_{e}}{3}-4(1-2x)uS_{2}^{\prime}\right)^2}{\left(\frac{\mu_{e}}{3x}+8 S_{2}\right)} 
\,.
\label{end-eq}
\end{eqnarray}
Comparing Eq.(\ref{end-eq}) with Eqs.~(\ref{dc22}), (\ref{nbdnbdu}) and (\ref{dmudx}) in the text, namely,
\begin{eqnarray}
c_s^2 &=&  
c_e^2 + \frac { \left[n_B \left( \frac {\partial \tilde \mu} {\partial n_B} \right)_x \right]^2} { \mu_{n}\left( \frac {\partial \tilde \mu}{\partial x} \right)_{n_B} } \\ 
n_B \frac {\partial \tilde \mu}{\partial n_B} &=& \frac {\mu_e}{3} - 4 (1-2x)~ u S_{2}^\prime \\
\frac {\partial \tilde \mu}{\partial x} &=& \frac 13 \frac {\mu_e}{x} + 8 S_2(u) \,,
\end{eqnarray}
we see that Eq.~(\ref{dc22}) is consistent with the direct definition of $c_s^2$ from Eq.~(\ref{cs2du}) and that Eq.~(\ref{dc22}) applies in general for any 
form of the EOS with a parabolic dependence in $x$, although in the text we arrived at Eqs.~(\ref{nbdnbdu}) and (\ref{dmudx}) in the context of the ZL-EOS.

\subsection{Derivation of Eqs.~(\ref{csmince})-(\ref{dydnmu})}

\noindent In $npe\mu$ matter, we choose the independent variables 
to be
the baryon density $n_B$, the lepton fraction $x$, and the muon fraction $y\equiv x_{\mu}$. The electron fraction $x_e$ is the difference $x-y$. \\ \\
The starting point for the speed-of-sound difference is 
\be
c_s^2 - c_e^2 = \frac{1}{\mu_{avg}}\left.\frac{\partial P}{\partial n_B}\right|_{x,y} 
                -\frac{1}{\mu_n}\frac{dP}{dn_B} ~.
\ee
The total derivative of the pressure $P(n_B,x,y)$ with respect to $n_B$ is given by
\be 
\frac{dP}{dn_B} = \left.\frac{\partial P}{\partial n_B}\right|_{x,y} 
                  +\left.\frac{\partial P}{\partial x}\right|_{n_B,y}\frac{dx}{dn_B}
                  +\left.\frac{\partial P}{\partial y}\right|_{n_B,x}\frac{dy}{dn_B}
\ee
and therefore
\beq
c_s^2 &-& c_e^2 = \left(\frac{1}{\mu_{avg}}-\frac{1}{\mu_n}\right) 
                  \left.\frac{\partial P}{\partial n_B}\right|_{x,y} \nonumber \\
               &-&\frac{1}{\mu_n}\left(\left.\frac{\partial P}{\partial x}\right|_{n_B,y}\frac{dx}{dn_B}
                  +\left.\frac{\partial P}{\partial y}\right|_{n_B,x}\frac{dy}{dn_B}\right)   \\
              &=& \left(\frac{\mu_n-\mu_{avg}}{\mu_{avg}~\mu_n}\right) 
                  \left.\frac{\partial P}{\partial n_B}\right|_{x,y}  \nonumber \\
                  &-&\frac{1}{\mu_n}\left(\left.\frac{\partial P}{\partial x}\right|_{n_B,y}\frac{dx}{dn_B}
                  +\left.\frac{\partial P}{\partial y}\right|_{n_B,x}\frac{dy}{dn_B}\right).   
\eeq
The average chemical potential $\mu_{avg}$ is 
\be
\mu_{avg} = (1-x)\mu_n + x\mu_p + (x-y)\mu_e + y\mu_y
\ee
which means that 
\beq
\mu_n - \mu_{avg} &=& x(\mu_n-\mu_p-\mu_e)+y(\mu_e-\mu_{\mu})    \nonumber \\
               &\equiv& -x \tilde\mu_x - y \tilde\mu_y  
\eeq
with the obvious definitions for $\tilde\mu_x$ and $\tilde\mu_y$.
\\ \\
In $\beta$-equilibrium $\mu_n = \mu_{avg}$ (as well as $\tilde\mu_x = \tilde\mu_y= 0$) and, correspondingly, 
\be
c_s^2 - c_e^2 =   -\frac{1}{\mu_n}\left(\left.\frac{\partial P}{\partial x}\right|_{n_B,y}\frac{dx}{dn_B}
                  +\left.\frac{\partial P}{\partial y}\right|_{n_B,x}\frac{dy}{dn_B}\right) ~, 
\ee
which is Eq.~(\ref{csmince}) in the main text.
Using $P = n_B^2 \left.\frac{\partial E}{\partial n_B}\right|_{x,y}$, the speed-of-sound difference can be expressed as
\beq
c_s^2 - c_e^2 &=& -\frac{n_B^2}{\mu_n}\frac{\partial}{\partial n_B}\left.
                   \left(\left.\frac{\partial E}{\partial x}\right|_{n_B,y}\frac{dx}{dn_B}
                  +\left.\frac{\partial E}{\partial y}\right|_{n_B,x}\frac{dy}{dn_B}\right)  
                  \right|_{x,y} \nonumber \\
              &=& -\frac{n_B^2}{\mu_n}
                  \left(\left.\frac{\partial \tilde\mu_x}{\partial n_B}\right|_{x,y}\frac{dx}{dn_B}
                  +\left.\frac{\partial \tilde\mu_y}{\partial n_B}\right|_{x,y}\frac{dy}{dn_B}\right)~.   \label{app_deltacs}
\eeq
The calculation of $\frac{dx}{dn_B}$ and $\frac{dy}{dn_B}$ begins from the total differentials of $\tilde\mu_x$ and $\tilde\mu_y$ which are 
\be
d\tilde\mu_x = \left.\frac{\partial \tilde\mu_x}{\partial n_B}\right|_{x,y}dn_B
             + \left.\frac{\partial \tilde\mu_x}{\partial x}\right|_{n_B,y}dx
             + \left.\frac{\partial \tilde\mu_x}{\partial y}\right|_{n_B,x}dy = 0
\ee
and
\be
d\tilde\mu_y = \left.\frac{\partial \tilde\mu_y}{\partial n_B}\right|_{x,y}dn_B
             + \left.\frac{\partial \tilde\mu_y}{\partial x}\right|_{n_B,y}dx
             + \left.\frac{\partial \tilde\mu_y}{\partial y}\right|_{n_B,x}dy = 0 ~.
\ee
From the former differential, it follows that 
\be
dy = \frac{-1}{\left.\frac{\partial \tilde\mu_x}{\partial y}\right|_{n_B,x}}
     \left(\left.\frac{\partial \tilde\mu_x}{\partial n_B}\right|_{x,y}dn_B
     + \left.\frac{\partial \tilde\mu_x}{\partial x}\right|_{n_B,y}dx \right)
\ee
which, when substituted in the latter, leads to
\beq
0 &=& \left.\frac{\partial \tilde\mu_y}{\partial n_B}\right|_{x,y}dn_B
             + \left.\frac{\partial \tilde\mu_y}{\partial x}\right|_{n_B,y}dx
             \nonumber \\
             &-& \frac{\left.\frac{\partial \tilde\mu_y}{\partial y}\right|_{n_B,x}}
                    {\left.\frac{\partial \tilde\mu_x}{\partial y}\right|_{n_B,x}}
     \left(\left.\frac{\partial \tilde\mu_x}{\partial n_B}\right|_{x,y}dn_B
             + \left.\frac{\partial \tilde\mu_x}{\partial x}\right|_{n_B,y}dx \right) ~.
\eeq
One then collects terms proportional to $dn_B$ and $dx$
\beq
&&\left(\left.\frac{\partial \tilde\mu_y}{\partial n_B}\right|_{x,y}
 -\frac{\left.\frac{\partial \tilde\mu_y}{\partial y}\right|_{n_B,x}}
       {\left.\frac{\partial \tilde\mu_x}{\partial y}\right|_{n_B,x}}
       \left.\frac{\partial \tilde\mu_x}{\partial n_B}\right|_{x,y} \right)dn_B \nonumber \\
       &=& -\left(\left.\frac{\partial \tilde\mu_y}{\partial x}\right|_{n_B,y}
 -\frac{\left.\frac{\partial \tilde\mu_y}{\partial y}\right|_{n_B,x}}
       {\left.\frac{\partial \tilde\mu_x}{\partial y}\right|_{n_B,x}}
       \left.\frac{\partial \tilde\mu_x}{\partial x}\right|_{n_B,y} \right)dx      
\eeq
or, equivalently,
\begin{equation}
\frac{dx}{dn_B} = \frac{\left.\frac{\partial \tilde\mu_x}{\partial y}\right|_{n_B,x}
                        \left.\frac{\partial \tilde\mu_y}{\partial n_B}\right|_{x,y}
                       -\left.\frac{\partial \tilde\mu_y}{\partial y}\right|_{n_B,x}
                       \left.\frac{\partial \tilde\mu_x}{\partial n_B}\right|_{x,y}}
                       {\left.\frac{\partial \tilde\mu_x}{\partial y}\right|_{n_B,x}
                       \left.\frac{\partial \tilde\mu_y}{\partial x}\right|_{n_B,y}
                       -\left.\frac{\partial \tilde\mu_y}{\partial y}\right|_{n_B,x}
                       \left.\frac{\partial \tilde\mu_x}{\partial x}\right|_{n_B,y}}  ~.
                       \label{app_dxdn}
\end{equation}
Similarly, 
\begin{equation}
\frac{dy}{dn_B} = \frac{\left.\frac{\partial \tilde\mu_x}{\partial x}\right|_{n_B,y}
                         \left.\frac{\partial \tilde\mu_y}{\partial n_B}\right|_{x,y}
                       -\left.\frac{\partial \tilde\mu_y}{\partial x}\right|_{n_B,y}
                       \left.\frac{\partial \tilde\mu_x}{\partial n_B}\right|_{x,y}}
                       {\left.\frac{\partial \tilde\mu_x}{\partial x}\right|_{n_B,y}
                       \left.\frac{\partial \tilde\mu_y}{\partial y}\right|_{n_B,x}
                       -\left.\frac{\partial \tilde\mu_y}{\partial x}\right|_{n_B,y}
                       \left.\frac{\partial \tilde\mu_x}{\partial y}\right|_{n_B,x}}  ~.
                       \label{app_dydn}
\end{equation}
\\
The speed-of-sound difference, as given by Eqs. (\ref{app_deltacs}), (\ref{app_dxdn}), and (\ref{app_dydn}) is physically transparent because $\beta$-equilibrium and compositional gradients are brought to the forefront via $\tilde\mu_{x(y)}$ and $\partial/\partial x(y)$, respectively (the latter two equations are 
Eqs.~(\ref{dxdnmu}) and (\ref{dydnmu}) in the main text). However, this intuitive picture comes at the expense of computational complexity.






\clearpage
\bibliography{apssamp}

\end{document}